\documentclass[3p,authoryear]{elsarticle}
\usepackage{epstopdf}
\usepackage{subfigure,graphicx,tabularx}
\usepackage{float}
\usepackage{amsmath, amsfonts, amssymb,mathrsfs}
\usepackage{amsthm}
\usepackage{epsf}
\usepackage{amsfonts}
\usepackage{stmaryrd}
\usepackage{amssymb}
\usepackage{leftidx}
\usepackage{color}
\usepackage{mathtools}
\usepackage{placeins}
\usepackage{booktabs}
\usepackage{enumitem}
\usepackage{caption}
\usepackage{multirow}
\usepackage{stackengine}
\usepackage[utf8]{inputenc}
\usepackage[english]{babel}
\usepackage{color}
\usepackage{pifont}
\usepackage{fontawesome}
\newcolumntype{?}{!{\vrule width 1pt}}

\theoremstyle{plain}
\newtheorem{theorem}{Theorem}

\newtheorem{remark}[theorem]{Remark}

\def\bfe{{\bf e}}

\def\bfs{{\bf s}}
\def\bfu{{\bf u}}

\def\bfx{{\bf x}}
\def\bfy{{\bf y}}

\def\bfE{{\bf E}}

\def\bfI{{\bf I}}

\def\bfN{{\bf N}}

\def\bfR{{\bf R}}
\def\bfS{{\bf S}}

\def\bfX{{\bf X}}
\def\bfF{{\bf F}}
\def\bfU{{\bf U}}

\newcommand\sts{s_{\texttt{ts}}}
\newcommand\scs{s_{\texttt{cs}}}
\newcommand\shs{s_{\texttt{hs}}}
\newcommand\sbs{s_{\texttt{bs}}}
\newcommand\sss{s_{\texttt{ss}}}

\long\def\symbolfootnote[#1]#2{\begingroup%
\def\thefootnote{\fnsymbol{footnote}}\footnote[#1]{#2}\endgroup}

\makeatletter
\renewcommand\@biblabel[1]{}

\makeatother

\begin{document}
\begin{frontmatter}

\title{Nine circles of elastic brittle fracture: A series of challenge problems to assess fracture models\vspace{0.1cm}}

\author[Illinois]{Farhad Kamarei}
\ead{kamarei2@illinois.edu}

\author[Duke]{Bo Zeng}
\ead{Bo.zeng@duke.edu}

\author[Duke]{John E. Dolbow}
\ead{jdolbow@duke.edu}

\author[Illinois]{Oscar Lopez-Pamies}
\ead{pamies@illinois.edu}

\address[Illinois]{Department of Civil and Environmental Engineering, University of Illinois, Urbana--Champaign, IL 61801, USA  \vspace{0.05cm}}

\address[Duke]{Department of Mechanical Engineering, Duke University, Durham, NC 27708, USA \vspace{0.05cm}}

\vspace{-0.1cm}

\begin{abstract}

Since the turn of the millennium, capitalizing on modern advances in mathematics and computation, a slew of computational models have been proposed in the literature with the objective of describing the nucleation and propagation of fracture in materials subjected to mechanical, thermal, and/or other types of loads. By and large, each new proposal focuses on a particular aspect of the problem, while ignoring others that have been well-established. This approach has resulted in a plethora of models that are, at best, descriptors of fracture only under a restricted set of conditions, while they may predict grossly incorrect and even non-physical behaviors in general. In an attempt to address this predicament, this paper introduces a vetting process in the form of nine challenge problems that any computational model of fracture must convincingly handle if it is to potentially describe fracture nucleation and propagation in general. The focus is on the most basic of settings, that of isotropic elastic brittle materials subjected to quasi-static mechanical loads. The challenge problems have been carefully selected so that: $i$) they can be carried out experimentally with standard testing equipment; $ii$) they can be unambiguously analyzed with a sharp description of fracture; and, most critically, $iii$) in aggregate they span the entire range of well settled experimental knowledge on fracture nucleation and propagation that has been amassed for over a century. For demonstration purposes, after their introduction, each challenge problem is solved with two phase-field models of fracture, a classical variational phase-field model and the phase-field model initiated by Kumar, Francfort, and Lopez-Pamies (\emph{J. Mech. Phys. Solids} 112 (2018), 523--551), this both for a prototypical elastic brittle hard material (soda-lime glass) and a prototypical elastic brittle soft material (a polyurethane elastomer).

\endkeyword

\end{abstract}

\end{frontmatter}

\vspace{-0.2cm}

\section{Introduction}\label{Sec: Intro}

Fueled by mathematical and computational progress at the turn of the millennium, numerous computational models have been and continue to be proposed in the literature to describe the phenomenon of fracture, that is, where and when cracks nucleate and propagate in materials subjected to external forces; see, e.g., \cite{Bourdin00,Silling00,Miehe10,Ortiz12,Iurlano16,Wu17,Anand18,KFLP18,KBFLP20,Weinberg19,Leguillon2020,Bobaru2021,Moes2022,Bazant23,Maurini24,Li2025}; and \cite{Lamen2025} among many others. Consciously or not, the majority of such models focus on describing a particular aspect of the problem, while ignoring others. Given the inherent complexity of fracture, in the best of cases, such models turn out to be descriptive of actual fracture only under limited conditions, while they may predict grossly incorrect and even non-physical behaviors in general; see, e.g., the recent review by \cite{LPDFL25} and \cite{KDLP25}.

In this context, the objective of this paper is to introduce a series of challenge problems --- or, put differently, an obstacle course\footnote{The idea of using an obstacle course to help assess the viability of computational models appears to date back to the work of \cite{Belytschko1985} in the context of shell finite elements.} --- to aid in assessing the viability of any proposed computational model of fracture. To keep the assessment as fundamental as possible, the problems are restricted to the most basic of settings, that of isotropic elastic brittle materials subjected to quasi-static mechanical loads. Each challenge problem has been carefully selected with three key characteristics in mind. First, they all can be readily performed in a laboratory using standard testing equipment. Second, they lend themselves to an unambiguous analysis through a sharp description of fracture. Most importantly, this collection of problems spans the complete spectrum of well-established experimental findings on both the nucleation and the propagation of fracture. Specifically, they probe fracture nucleation governed by the strength of the material, when the material is subjected to spatially uniform stresses. They probe fracture nucleation governed by the Griffith competition between bulk deformation and surface fracture energies, when nucleation occurs from the front of a large pre-existing crack. They probe fracture nucleation governed by the mediation between strength and Griffith, when nucleation occurs in regions where the stresses are not spatially uniform. Finally, they probe fracture propagation governed by the Griffith energy competition, both in opening (Mode I) and tearing (Mode III) modes. In total, there are nine challenge problems.\footnote{The number of problems being the same as the number of circles in Dante's Inferno may not be fortuitous.}  They are listed in Table \ref{Table0}, alongside the type of fracture nucleation and/or propagation that they characterize. If a model fails to deliver accurate predictions for one of these problems, then such a model is \emph{not} a viable candidate to describe --- and hence predict --- fracture in general.  
\begin{table}[t!]\centering
\caption{The nine challenge problems and the type of fracture nucleation and/or propagation that they characterize.}
\begin{tabular}{r|ccccc}
\toprule
                              & Strength      &    Griffith       & Strength-Griffith    & Griffith       & Griffith  \\
                              & Nucleation    &    Nucleation     & Mediated Nucleation  & Propagation    & Propagation  \\
                              &               &                   &                      & Mode I         & Mode III  \\
\midrule
Uniaxial tension             & $\checkmark$   &                   &                      &                  &           \\
\midrule
Biaxial tension              & $\checkmark$   &                   &                      &                  &           \\
\midrule
Torsion                       & $\checkmark$   &                   &                      &                  &           \\
\midrule
Pure-shear                   &                &    $\checkmark$    &                      &                 &           \\
\midrule
Single edge notch            &                &                    &     $\checkmark$     &                 &            \\
\midrule
Indentation                  &                &                    &     $\checkmark$     &                 &            \\
\midrule
Poker-chip                   &                &                    &     $\checkmark$     &                 &            \\
\midrule
Double cantilever beam       &                &    $\checkmark$    &                      &  $\checkmark$   &            \\
\midrule
Trousers                     &                &    $\checkmark$    &                      &                 & $\checkmark$  \\
\bottomrule
\end{tabular} \label{Table0}
\end{table}

To illustrate their deployment and utility, we solve each challenge problem with two distinct phase-field models: the classical variational \texttt{AT}$_1$ formulation, originally employed by \cite{Marigo11} and later popularized by \cite{Tanne18}, and the formulation initiated by \cite*{KFLP18}. Furthermore, we make use of two distinct open-source finite-element (FE) codes to solve them, one being the platform FEniCS, the other being RACCOON (within MOOSE). Both of these codes, together with the corresponding FE meshes for all nine problems, have been made available on GitHub.\footnote{https://github.com/farhadkama/FEniCSx\_Kamarei\_Lopez-Pamies.}$^,$\footnote{https://github.com/hugary1995/raccoon.} 

For completeness, the proposed challenge problems are solved for a soda-lime glass, a prototypical hard material, as well as for a polyurethane (PU) elastomer, a prototypical soft material. Consistent with classical experimental results \citep{Guin19,Meyland21}, the soda-lime glass is taken to be an isotropic linear elastic brittle material with elastic energy density
\begin{equation}
W(\bfE)=\mu\, {\rm tr}\,\bfE^2+\dfrac{\Lambda}{2}({\rm tr}\,\bfE)^2,\label{W-Lin}
\end{equation}
Drucker-Prager strength surface 
\begin{equation}
\mathcal{F}(\bfS)=\sqrt{\mathcal{J}_2}+\dfrac{s_{\texttt{ts}}}
{\sqrt{3}\left(3 s_{\texttt{hs}}-s_{\texttt{ts}}\right)}\, \mathcal{I}_1-\dfrac{\sqrt{3}s_{\texttt{hs}} s_{\texttt{ts}}}
{3s_{\texttt{hs}}-s_{\texttt{ts}}}=0,\label{DP}
\end{equation}
scalar critical energy release rate
\begin{equation}
G_c,\label{Gc}
\end{equation}
and the material constants listed in Table \ref{Table1}. On the other hand, the PU elastomer is taken to be an isotropic non-linear elastic brittle material with Neo-Hookean elastic energy density
\begin{equation}
\psi(\bfF)=\dfrac{\mu}{2}\left({\rm tr}\,(\bfF^T\bfF)-3\right)-\mu\ln(\det\bfF)+\dfrac{\Lambda}{2}(\det\bfF-1)^2,\label{W-NH}
\end{equation}
Drucker-Prager strength surface (\ref{DP}), scalar critical energy release rate (\ref{Gc}), and the material constants listed in Table \ref{Table2}; this latter description pertains to one of the PU elastomers with short polymer chains, lack of entanglements, and hence highly elastic behavior studied by \cite{Creton10}. In the above expressions, $\bfE=\frac{1}{2}(\nabla\bfu+\nabla\bfu^T)$ and $\bfF=\bfI+\nabla\bfu$ stand for the infinitesimal strain tensor and the deformation gradient tensor, with $\bfu$ denoting the displacement field, while $\bfS$ is the first Piola-Kirchhoff stress tensor, $\mathcal{I}_1=s_1+s_2+s_3$, $\mathcal{J}_2=\frac{1}{3}(s_1^2+s_2^2+s_3^2-s_1 s_2-s_1 s_3-s_2 s_3)$,  and $s_1$, $s_2$, $s_3$ stand for the eigenvalues of the Biot stress tensor, or principal nominal stresses. Furthermore, $\mu$ and $\Lambda$ stand for the initial shear modulus and first Lam\'e constant,  while $\sts$ and $\shs$ denote the nominal uniaxial tensile and hydrostatic strengths. For later reference, the corresponding values for the bulk modulus $\kappa=\Lambda+2\mu/3$, the Young's modulus $E=(3\Lambda+2\mu)\mu/(\Lambda+\mu)$, the Poisson's ratio $\nu=\Lambda/(2(\Lambda+\mu))$, the biaxial tensile strength $\sbs=3\shs\sts/(3\shs+\sts)$, the shear strength $\sss=\sqrt{3}\shs\sts/(3\shs-\sts)$, and the uniaxial compressive strength $\scs=3\shs\sts/(3\shs-2\sts)$ are also listed in Tables \ref{Table1} and \ref{Table2}.
\begin{table}[H]\centering
\caption{Material constants for the soda-lime glass used as a prototypical hard material in the challenge problems.}
\begin{tabular}{l?cc|ccc}
\toprule
Elasticity constants & $\mu$ (GPa)& $\Lambda$ (GPa) & $\kappa$ (GPa)  & $E$ (GPa)  & $\nu$ \\
\midrule
                     & $28.7$   & $22.5$        & $41.7$         & $70$   & $0.22$ \\
\midrule
\midrule
Strength constants   & $s_{\texttt{ts}}$ (MPa) & $s_{\texttt{hs}}$ (MPa) & $s_{\texttt{bs}}$ (MPa)  & $s_{\texttt{ss}}$ (MPa)  & $s_{\texttt{cs}}$ (MPa)\\
\midrule
                     & $40$                  & $27.8$                  & $27$           & 44.4   & 1000 \\
\midrule
\midrule
Critical energy release rate   & $G_c$ (N/m) &   &  &  &  \\
\midrule
                               & $10$ & &  &  & \\
\bottomrule
\end{tabular} \label{Table1}
\end{table}
\begin{table}[H]\centering
\caption{Material constants for the PU elastomer used as a prototypical soft material in the challenge problems.}
\begin{tabular}{l?cc|cc}
\toprule
Elasticity constants & $\mu$ (MPa)& $\Lambda$ (MPa) &   &   \\
\midrule
                     & $0.52$   & $85.77$        &          &    \\
\midrule
\midrule
Strength constants   & $s_{\texttt{ts}}$ (MPa) & $s_{\texttt{hs}}$ (MPa) & $s_{\texttt{bs}}$ (MPa)  & $s_{\texttt{ss}}$ (MPa)   \\
\midrule
                     & $0.3$                  & $1$                  & $0.27$           & 0.19   \\
\midrule
\midrule
Critical energy release rate   & $G_c$ (N/m) &   &  &  \\
\midrule
                               & $41$ & &  &  \\
\bottomrule
\end{tabular} \label{Table2}
\end{table}

The paper is organized as follows. In Sections \ref{Sec: Strength Nucleation}, \ref{Sec: Griffith Nucleation}, and \ref{Sec: Mediation Nucleation}, we introduce, analyze, and discuss the challenge problems that characterize fracture nucleation governed by the material strength, by the Griffith energy competition, and by the mediation between these two properties. In Section \ref{Sec: Griffith Propagation} we do the same for the challenge problems that characterize fracture propagation. We close by recording a number of final comments in Section \ref{Sec: Final comments}. Appendices A and B provide summaries of both the phase-field model by \cite*{KFLP18} and the classical variational \texttt{AT}$_1$ formulation, the latter being presented as a special case of the former.



\section{Fracture nucleation governed by strength}\label{Sec: Strength Nucleation}

Historically, the first investigations of fracture centered on uniaxial tension tests, where a gauge section in a specimen of the material of interest was subjected to a spatially uniform uniaxial tensile stress $\bfS={\rm diag}(s>0,0,0)$ until a crack suddenly appeared severing the specimen; see, e.g., \cite{Lame1833}, Section 83 in the monograph by \cite{Love1906}, and \cite{Busse34}. The critical value $\sts$ of the nominal stress $s$ at which the specimen fractured identified the uniaxial tensile strength of the material. By now, it has been well established that the set of all critical stresses at which a material fractures when it is subjected to a state of monotonically increasing, \emph{spatially uniform}, but otherwise arbitrary stress defines the strength of that material in its entirety.\footnote{Despite the fact that it was the study of uniaxial tensile strength that, in earnest, started research into fracture over almost two centuries ago, the precise and complete definition of strength stated here for general stress states was only introduced a few years ago \citep{KLP20,KBFLP20}. As will become apparent below, such a prolonged misunderstanding and mistreatment of strength as an intrinsic macroscopic material property has been one of the main reasons why the majority of computational models of fracture that have been developed over the years are \emph{not} descriptive of actual fracture.} Such a set of critical stresses defines a surface 
$$\mathcal{F}(\bfS)=0$$
in stress space, which is referred to as the \emph{strength surface} of the material.\footnote{Precisely, $\mathcal{F}(\bfS)=0$ is potentially any star-shaped --- and thus possibly non-convex --- surface in stress space containing ${\bf 0}$ in its interior. Accordingly, rays starting at the origin $\bfS={\bf 0}$ can cross the strength surface $\mathcal{F}(\bfS)=0$ at most once.} 

In practice, it is difficult to carry out experiments that probe the entire strength surface $\mathcal{F}(\bfS)=0$ of any given material, be it hard or soft, but a good number of tests have been developed over the years that allow to probe a range of roughly spatially uniform biaxial states of stress  $\bfS={\rm diag}(s_1,s_2,0)$ right up to fracture nucleation. These tests include the combined axial loading, pressurization, and torsion of thin-walled tubes, the biaxial loading of plates and sheets of various shapes, and the inflation or bulging of thin films through openings of various shapes; see, e.g., \cite{Treloar1947,Knauss67,Ely72,Kawabata1973,Sato87,Kim92,Sasso2008}, and \cite{Ferraris2015}. 

Out of all the tests that have been developed to probe the strength of materials, the most robust are arguably uniaxial tension tests wherein the gauge section is a circular rod, biaxial tension tests wherein the gauge section is a circular plate, and torsion tests wherein the gauge section is a thin-walled circular tube. Such gauge sections are conducive to the development of fairly uniform states of stress of the form $\bfS={\rm diag}(s>0,0,0)$, $\bfS={\rm diag}(s>0,s>0,0)$,  and $\bfS={\rm diag}(s,-s,0)$, respectively. Moreover, the lack of corners in such gauge sections aids in minimizing the presence of specimen surface defects that may lead to specimen failures that are not representative of the actual strength of the material. For these reasons, together with the fact that they probe different archetypal states of stress, we choose these three tests as the challenge problems that any viable computational model of fracture must convincingly handle in its characterization of fracture nucleation governed by strength. In the following three subsections, we introduce, analyze, and discuss these three challenge problems in full detail, one at a time.

\subsection{Uniaxial tension test}

\begin{figure}[t!]
\centering
\centering\includegraphics[width=0.90\linewidth]{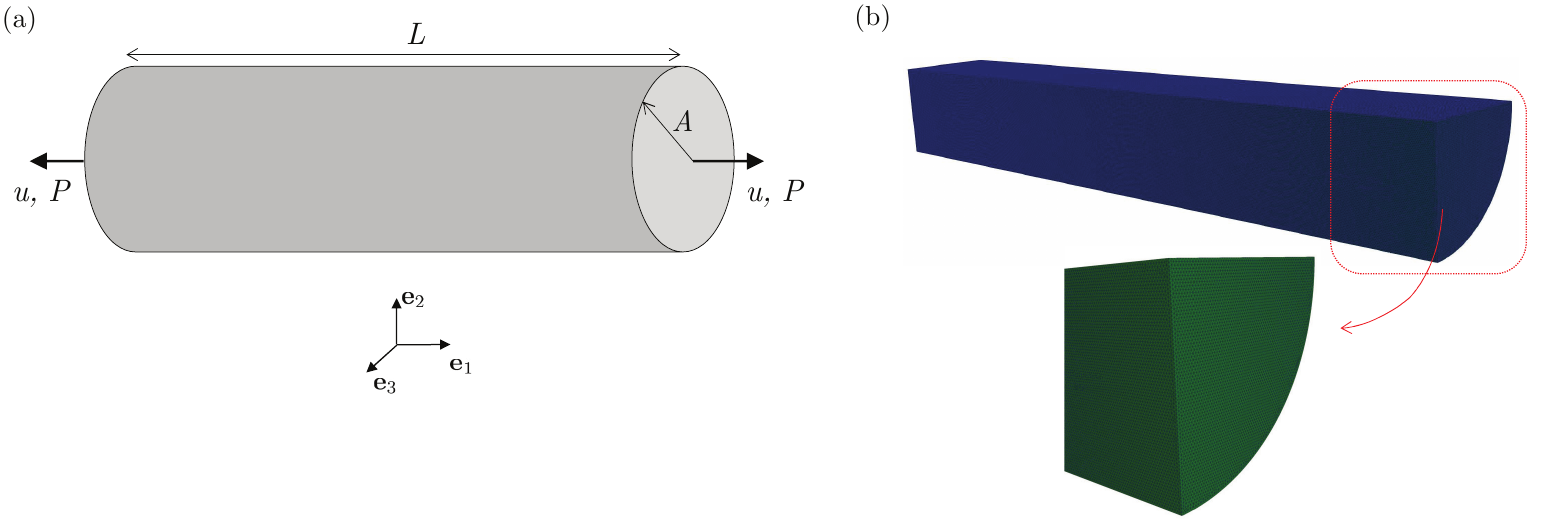}
\caption{\small (a) Schematic of the uniaxial tension test. The specimen dimensions are $L=15$ mm and $A=2$ mm. (b) Unstructured FE mesh of uniform element size $\texttt{h}=0.03$ mm utilized for the phase-field simulations of the test.}\label{Fig1}
\end{figure}

The first challenge problem, shown schematically in Fig.~\ref{Fig1}(a), is that of a circular rod, of initial length $L=15$ mm and radius $A=2$ mm, that is subjected to uniaxial tension at its ends by the application of an axial displacement $u$. The rod, which can be viewed as the gauge section in a larger specimen, undergoes a uniform axial strain $2u/L$ and corresponding uniform uniaxial stress $S=P/(\pi A^2)$, where $P$ is the resultant force at the ends of the rod. Precisely, in the Cartesian laboratory frame of reference indicated in Fig.~\ref{Fig1}(a),
\begin{equation*}
\left\{\hspace{-0.15cm}\begin{array}{l}
\bfE=\dfrac{2u}{L}\bfe_1\otimes\bfe_1+(\lambda_{t}-1)(\bfe_2\otimes\bfe_2+\bfe_3\otimes\bfe_3)\vspace{0.2cm}\\
\bfF=\left(1+\dfrac{2u}{L}\right)\bfe_1\otimes\bfe_1+\lambda_{t}(\bfe_2\otimes\bfe_2+\bfe_3\otimes\bfe_3)\end{array}\right.\quad {\rm and}\quad \bfS=S\bfe_1\otimes\bfe_1,
\end{equation*}
where $\lambda_t$ is the transverse stretch due to the Poisson effect. This state of spatially uniform stress and strain persists until the value of the stress $S$ reaches the uniaxial tensile strength $\sts$ of the material, at which point the rod is severed into two pieces by the sudden nucleation of a crack orthogonal to the direction of the applied displacement. The location where the crack nucleates is arbitrary as a direct consequence of the strength of the material being an inherently stochastic macroscopic material property. 

We now focus on specific results for the case when the rod is made of the soda-lime glass with the material constants listed in Table \ref{Table1}, as well as for the case when the rod is made of the PU elastomer with the material constants listed in Table \ref{Table2}. For these materials, the measures of stress $S$ and strain $2u/L$ are related according to
\begin{equation}
S=\left\{\hspace{-0.15cm}\begin{array}{ll}
E\dfrac{2u}{L}& {\rm if}\;\; 0\leq u<u_{\texttt{ts}}\vspace{0.2cm}\\
0&{\rm if} \;\;u_{\texttt{ts}}\leq u\end{array}\right. \;{\rm and}\; S=\left\{\hspace{-0.15cm}\begin{array}{ll}
\mu\left(1+\dfrac{2u}{L}-\dfrac{L^2}{(L+2u)^2}\right)-\dfrac{2 \mu^2 L^2 u}{\Lambda(L+2u)^3}+O(\Lambda^{-2})& {\rm if}\;\; 0\leq u<u_{\texttt{ts}}\vspace{0.2cm}\\
0&{\rm if} \;\;u_{\texttt{ts}}\leq u\end{array}\right.,\label{Response-Uni}
\end{equation}
respectively, where $u_{\texttt{ts}}$ denotes the value of the displacement at which $S=\sts$. For a computational model of fracture to be viable, it must be able to deliver predictions that agree with these results, as well as with the orientation of the nucleated crack being orthogonal to the direction of the applied displacement. 
\begin{figure}[b!]
\centering
\centering\includegraphics[width=0.99\linewidth]{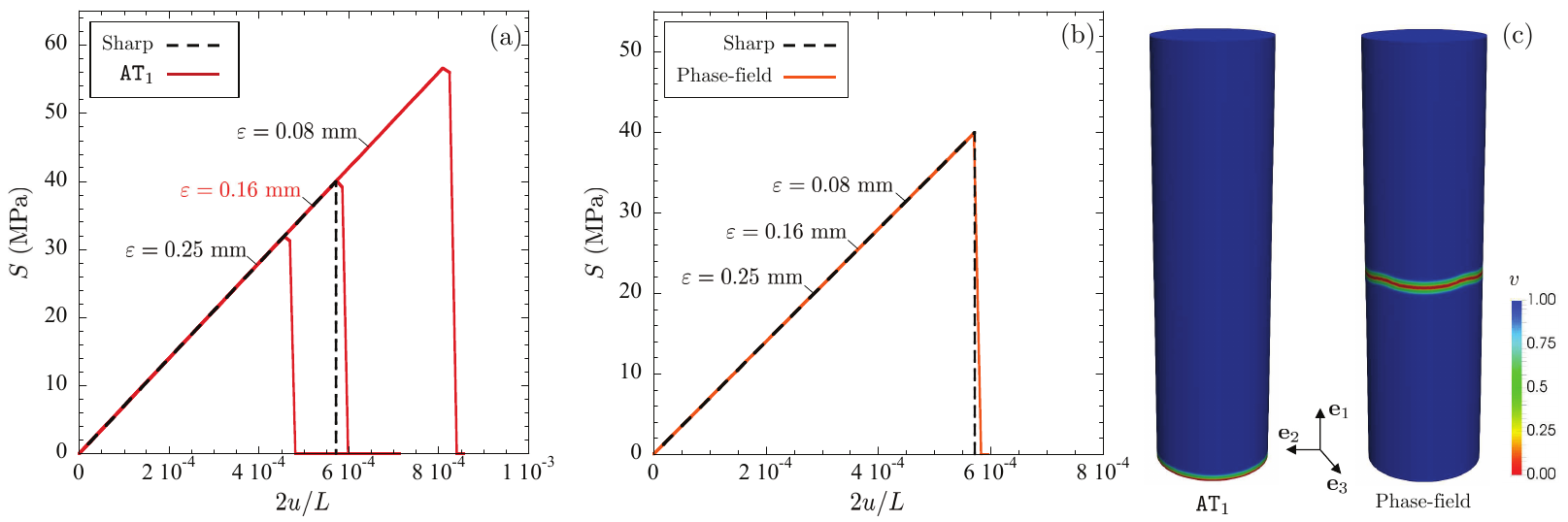}
\caption{\small Uniaxial tension test of a soda-lime glass rod. Comparisons between the exact result (\ref{Response-Uni})$_1$ for the stress-strain response of the rod and the predictions by (a) the \texttt{AT}$_1$ and (b) the phase-field models for three different values of the regularization length $\varepsilon$ in these models. (c) Contour plots of the phase field $v$ over the undeformed configuration of the rod right after fracture nucleation, as predicted by the \texttt{AT}$_1$ and phase-field models for $\varepsilon=0.16$ mm.}\label{Fig2}
\end{figure}

For demonstration purposes, as announced in the Introduction, we make use of the classical variational \texttt{AT}$_1$ phase-field model and of the phase-field model of \cite*{KFLP18} to simulate this and the subsequent challenge problems. For short, henceforth, we shall refer to the former as \emph{the} \texttt{AT}$_1$ \emph{model} and to the latter as \emph{the phase-field model}. Recall that the governing equations for these models, together with pertinent references, are summarized below, in Appendices A and B, and that these equations are solved numerically by means of the FE method, as implemented in the open-source platforms FEniCS and RACCOON. To reduce computational cost, we exploit symmetry and perform the simulations over a quarter of the rod; of course, one can exploit symmetry further and recast the problem as a 2D axisymmetric problem, but we prefer to stay within a 3D setting here to thoroughly test the performance of the models. Figure \ref{Fig1}(b) shows the corresponding FE mesh used to carry out the simulations. The mesh is unstructured and of uniform element size $\texttt{h}=0.03$ mm, which is sufficiently small to lead to converged solutions. For this and subsequent simulations, standard linear tetrahedral elements are used to generate solutions for the soda-lime glass, while non-conforming linear Crouzeix-Raviart elements are employed for the PU elastomer to address its near incompressibility ($\Lambda/\mu=165$). As RACCOON does not yet have the latter type of elements built in, results for this and the other challenge problems on the PU elastomer are generated only in FEniCS.

\begin{figure}[t]
\centering
\centering\includegraphics[width=0.99\linewidth]{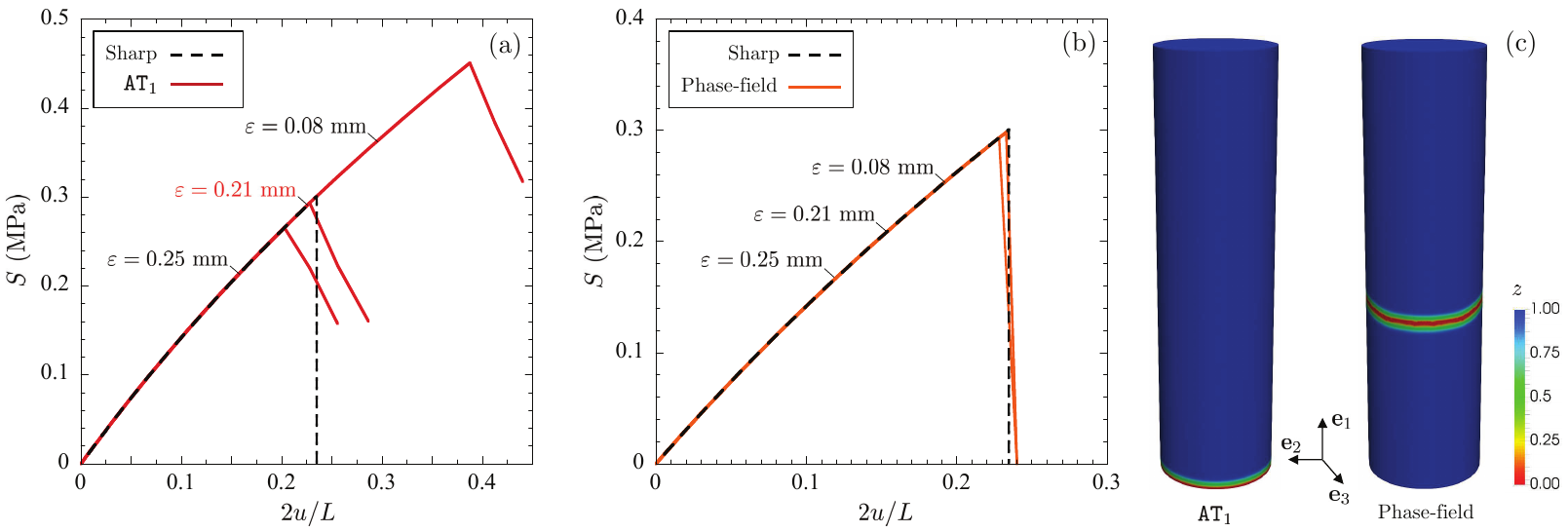}
\caption{\small Uniaxial tension test of a PU elastomer rod. Comparisons between the exact result (\ref{Response-Uni})$_2$ for the stress-strain response of the rod and the predictions by (a) the \texttt{AT}$_1$ and (b) the phase-field models for three different values of the regularization length $\varepsilon$. (c) Contour plots of the phase field $z$ over the undeformed configuration of the rod right after fracture nucleation, as predicted by the \texttt{AT}$_1$ and phase-field models for $\varepsilon=0.21$ mm.}\label{Fig3}
\end{figure}

Figures \ref{Fig2} and \ref{Fig3}  compare the predictions generated by the \texttt{AT}$_1$ and phase-field models with the exact results (\ref{Response-Uni}) --- labeled \emph{sharp} in the plots for their sharp description of fracture --- for the uniaxial tension tests on the soda-lime glass and PU elastomer. Specifically, parts (a) of the figures present comparisons for the stress-strain response ($S$ vs. $2u/L$) between the predictions by the \texttt{AT}$_1$ model and the exact results (\ref{Response-Uni}) for three values of the regularization length $\varepsilon$ of that model. Parts (b) present the same type of comparisons for the phase-field model. Finally, parts (c) present contour plots of the phase fields, $v$ and $z$, over the undeformed configuration of the rods, right after fracture nucleation has occurred, as predicted by the \texttt{AT}$_1$ and phase-field models for one of the values of the regularization length, $\varepsilon=0.16$ mm in Fig.~\ref{Fig2}(c) and $\varepsilon=0.21$ mm in Fig.~\ref{Fig3}(c).

There are two main observations from Figs.~\ref{Fig2} and \ref{Fig3}. First, the \texttt{AT}$_1$ model predicts the correct elastic deformation of the rods and also the approximately correct abrupt nucleation of a crack with the right orientation. Nevertheless, the critical value of the stress $S$ at which this crack nucleates, say $\sts^{\texttt{AT}_1}$, depends on the value of the regularization length $\varepsilon$. This is because, as expounded elsewhere \citep{KBFLP20,LPDFL25,KDLP25}, the \texttt{AT}$_1$ model does \emph{not} account for the actual strength surface $\mathcal{F}(\bfS)=0$ of the material, instead, it features a built-in strength surface $\mathcal{F}^{\texttt{AT}_1}(\bfS)=0$, defined in terms of the elastic energy density ($W(\bfE)$ or $\psi(\bfF)$) and the critical energy release rate ($G_c$) of the material, that depends on $\varepsilon$. For the uniaxial tension test of interest here, this built-in strength surface predicts the uniaxial tensile strength \citep{KDLP25}
\begin{equation*}
\sts^{\texttt{AT}_1}=\sqrt{\dfrac{3 G_c E}{8\varepsilon}}\quad {\rm and}\quad \sts^{\texttt{AT}_1}=\mu\left(\dfrac{l^3-1}{l^2}+\dfrac{\mu(1-l)}{\Lambda l^3}\right),
\end{equation*}
where $l$ is the root closest to $1$ of the non-linear algebraic equation $f_{\texttt{ts}}^{\texttt{AT}_1}(l;\varepsilon)=\mu(l^2+2/l-3-\mu(l-1)^2/(\Lambda l^2))-3G_c/(8\varepsilon)=0$, for the soda-lime glass and the PU elastomer, respectively. As illustrated by Figs.~\ref{Fig2}(a) and \ref{Fig3}(a), these predictions are such that $\sts^{\texttt{AT}_1}\nearrow +\infty$ as $\varepsilon\searrow 0$. They are also such that a particular value of $\varepsilon$ can be selected to match the actual uniaxial tensile strength $\sts$ of the material. For the soda-lime glass, this value is about $\varepsilon=0.16$ mm, while for the PU elastomer it is about $\varepsilon=0.21$ mm; these fitted values are written in red in the plots. Thus, the \texttt{AT}$_1$ model can deliver predictions that are in agreement with the nucleation of fracture in a uniaxial tension test provided that the value of the regularization length $\varepsilon$ is suitably selected. As demonstrated by the comparisons with the other challenge problems that follow --- contrary to a widespread misplaced belief in the literature --- this is a mere fitting exercise void of physical justification that cannot resolve the innate limitation by the \texttt{AT}$_1$ model of not accounting for the actual strength surface $\mathcal{F}(\bfS)=0$ of the material. 

The second main observation from Figs.~\ref{Fig2} and \ref{Fig3} is that, irrespective of the value of $\varepsilon$, the phase-field model predicts accurately  the exact results (\ref{Response-Uni}) in their entirety, as well as the associated correct abrupt nucleation of a crack with the right orientation.

\subsection{Biaxial tension test}

\begin{figure}[b!]
\centering
\centering\includegraphics[width=0.90\linewidth]{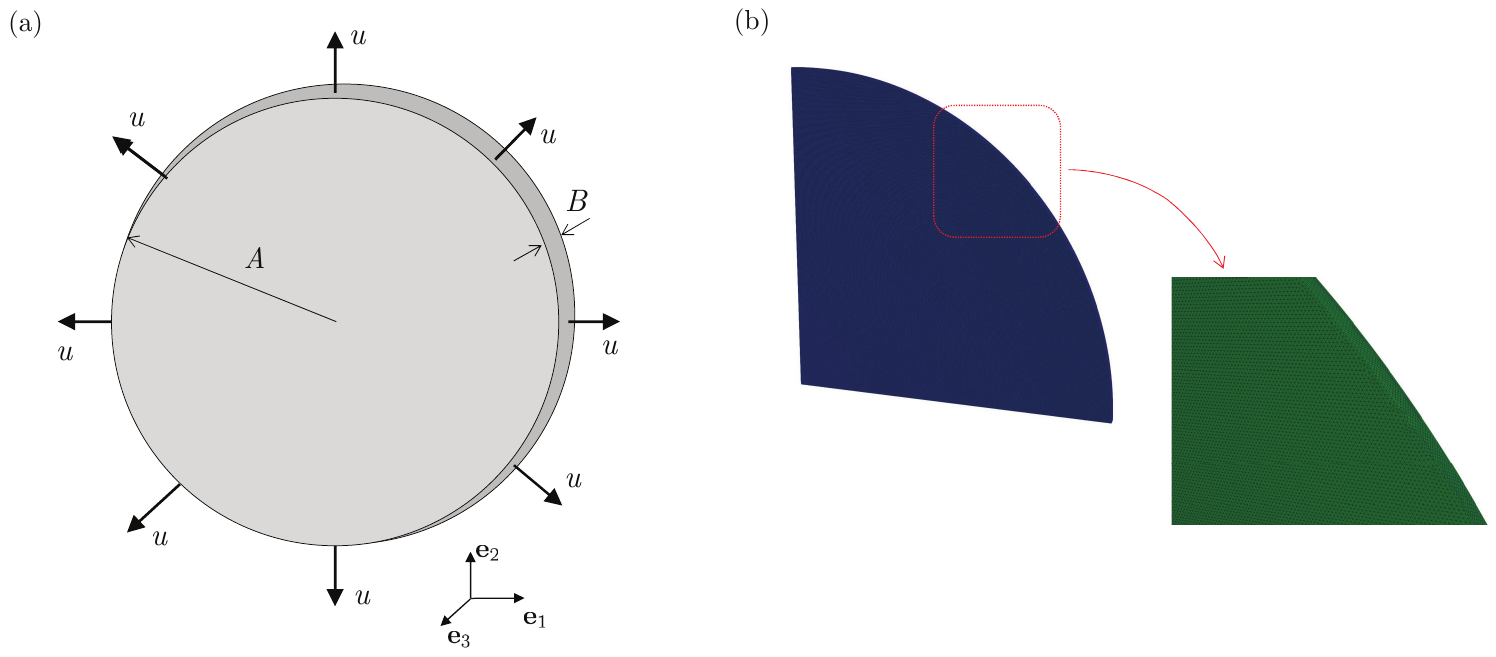}
\caption{\small (a) Schematic of the biaxial tension test. The specimen dimensions are $A=5$ mm and $B=0.25$ mm. (b) Unstructured FE mesh of uniform element size $\texttt{h}=0.015$ mm utilized for the phase-field simulations of the test.}\label{Fig4}
\end{figure}

The second challenge problem is depicted in Fig.~\ref{Fig4}(a). It involves a circular plate of radius $A=5$ mm and thickness $B=0.25$ mm, to be viewed as the gauge section in a larger specimen, that is subjected to (equi-)biaxial tension via an affine displacement $u$ applied at its lateral boundary $\partial\Omega_l$. Precisely, $\bfu=(u/A(\bfe_1\otimes\bfe_1+\bfe_2\otimes\bfe_2)+(\lambda_t-1)\bfe_3\otimes\bfe_3)\bfX$, $\bfX\in\partial\Omega_l=\{\bfX:X_1^2+X_2^2=A^2,\,-B/2<X_3<B/2\}$, where $\{\bfe_1,\bfe_2,\bfe_3\}$ stands for the Cartesian laboratory frame of reference indicated in the figure. This boundary condition results in a spatially uniform biaxial strain $u/A$ and corresponding uniform biaxial stress $S$ within the plate, namely, 
\begin{equation*}
\left\{\hspace{-0.15cm}\begin{array}{l}
\bfE=\dfrac{u}{A}(\bfe_1\otimes\bfe_1+\bfe_2\otimes\bfe_2)+(\lambda_t-1)\bfe_3\otimes\bfe_3\vspace{0.2cm}\\
\bfF=\left(1+\dfrac{u}{A}\right)(\bfe_1\otimes\bfe_1+\bfe_2\otimes\bfe_2)+\lambda_{t}\bfe_3\otimes\bfe_3\end{array}\right.\quad {\rm and}\quad \bfS=S(\bfe_1\otimes\bfe_1+\bfe_2\otimes\bfe_2),
\end{equation*}
where $\lambda_t$ is the transverse stretch due to the Poisson effect. This state of stress and strain continues until the value of the stress $S$ reaches the biaxial tensile strength $\sbs$ of the material. At that point, a through-thickness crack suddenly nucleates severing the plate into different pieces. In view of the stochastic nature of the strength of the material, the location where the crack nucleates is arbitrary.

For the cases when the plate is made of the soda-lime glass and PU elastomer,  the measures of stress $S$ and strain $u/A$ are related according to
\begin{equation}
S=\left\{\hspace{-0.15cm}\begin{array}{ll}
\dfrac{E u}{(1-\nu)A}& {\rm if}\;\; 0\leq u<u_{\texttt{bs}}\vspace{0.2cm}\\
0&{\rm if} \;\;u_{\texttt{bs}}\leq u\end{array}\right. \label{Response-Bi-Lin}
\end{equation}
and
\begin{equation}
S=\left\{\hspace{-0.15cm}\begin{array}{ll}
\mu\left(1+\dfrac{u}{A}-\dfrac{A^5}{(A+u)^5}\right)-\dfrac{2 \mu^2 A^5(A^4-(A+u)^4)}{\Lambda (A+u)^9}+O(\Lambda^{-2})& {\rm if}\;\; 0\leq u<u_{\texttt{bs}}\vspace{0.2cm}\\
0&{\rm if} \;\;u_{\texttt{bs}}\leq u\end{array}\right.,\label{Response-Bi-NH}
\end{equation}
respectively, where $u_{\texttt{bs}}$ denotes the value of the displacement at which $S=\sbs$. Again, for a computational model of fracture to be viable, it must be able to deliver predictions that agree with these results, in addition to being able to deliver predictions that agree with the previous results (\ref{Response-Uni}) for uniaxial tension.

\begin{figure}[b!]
\centering
\centering\includegraphics[width=0.99\linewidth]{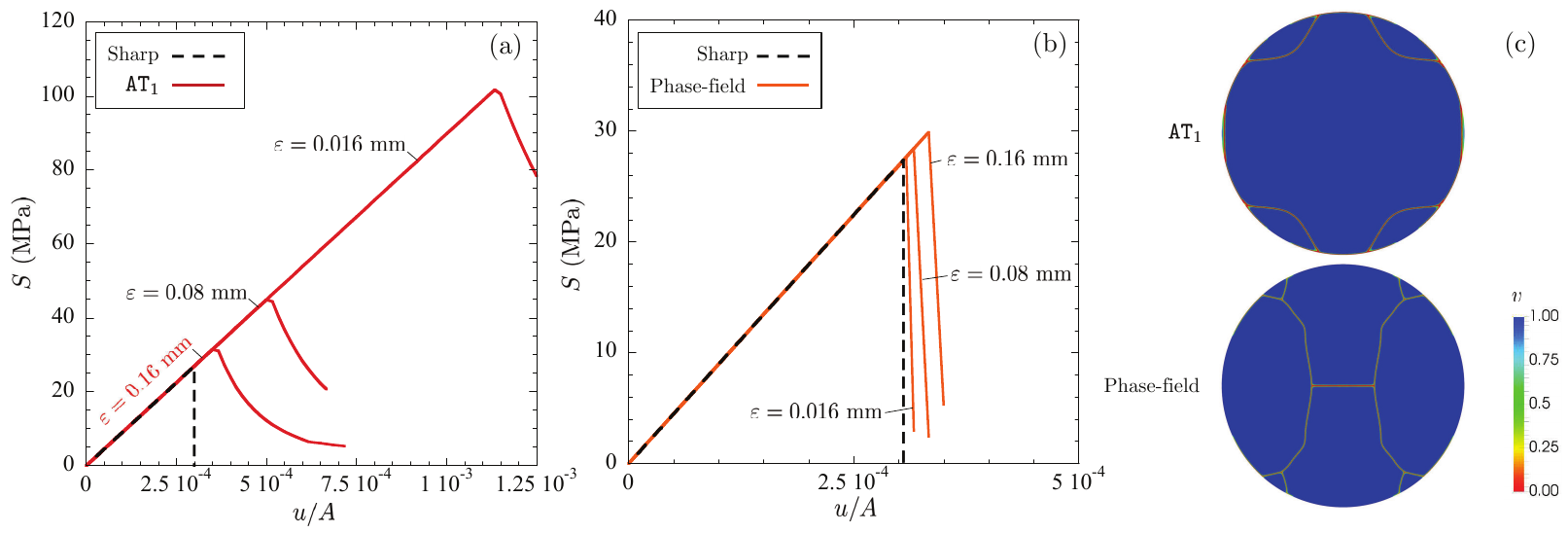}
\caption{\small Biaxial tension test of a soda-lime glass plate. Comparisons between the exact result (\ref{Response-Bi-Lin}) for the stress-strain response of the plate and the predictions by (a) the \texttt{AT}$_1$ and (b) the phase-field models for three different values of the regularization length $\varepsilon$. (c) Contour plots of the phase field $v$ over the undeformed configuration of the plate right after fracture nucleation, as predicted by the \texttt{AT}$_1$ and phase-field models for $\varepsilon=0.016$ mm.}\label{Fig5}
\end{figure}
\begin{figure}[t!]
\centering
\centering\includegraphics[width=0.99\linewidth]{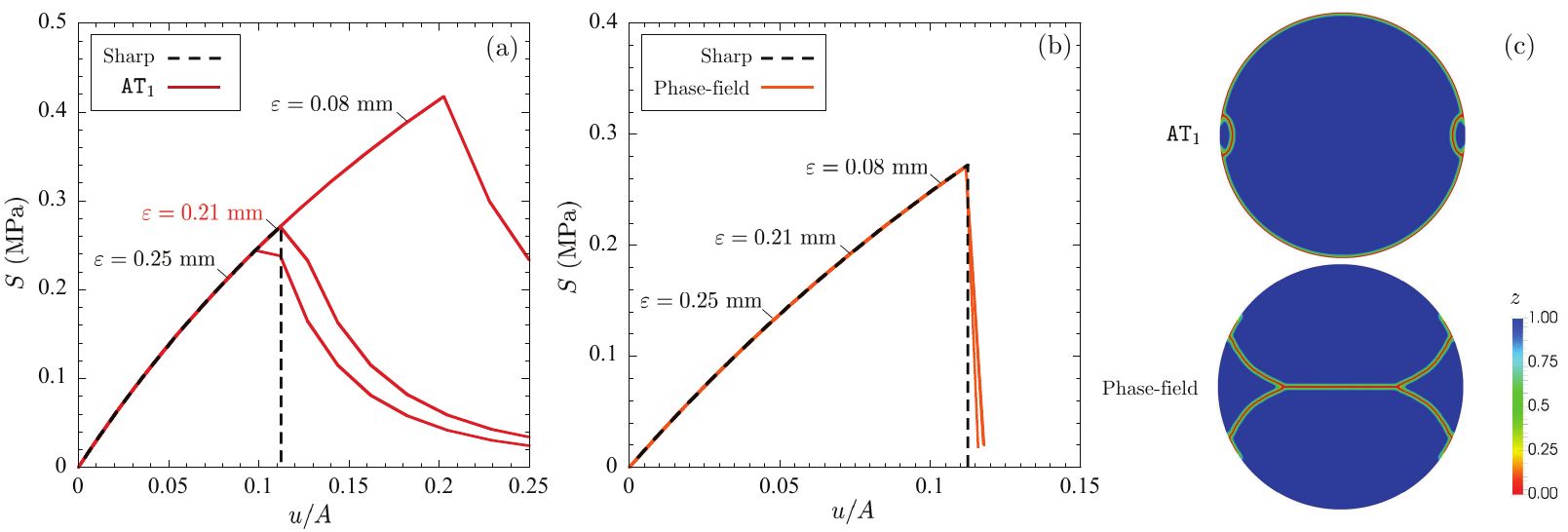}
\caption{\small Biaxial tension test of a PU elastomer plate. Comparisons between the exact result (\ref{Response-Bi-NH}) for the stress-strain response of the plate and the predictions by (a) the \texttt{AT}$_1$ and (b) the phase-field models for three different values of the regularization length $\varepsilon$. (c) Contour plots of the phase field $z$ over the undeformed configuration of the plate right after fracture nucleation, as predicted by the \texttt{AT}$_1$ and phase-field models for $\varepsilon=0.08$ mm.}\label{Fig6}
\end{figure}

In the simulations of the biaxial test by means of the \texttt{AT}$_1$ and phase-field models, to reduce computational cost, we exploit symmetry and perform them over an octant of the plate. 
The FE mesh used for these simulations is shown in Fig.~\ref{Fig4}(b). It is unstructured and of uniform element size $\texttt{h}=0.015$ mm, which is sufficiently small to lead to converged solutions. Figures \ref{Fig5} and \ref{Fig6} compare the predictions generated by the \texttt{AT}$_1$ and phase-field models with the exact results (\ref{Response-Bi-Lin}) and (\ref{Response-Bi-NH}) for the biaxial tension tests on the soda-lime glass and PU elastomer, respectively. Parts (a) present comparisons for the \texttt{AT}$_1$ model for three values of the regularization length $\varepsilon$, while parts (b) show the same type of comparisons for the phase-field model. Parts (c) provide contour plots of the phase fields, $v$ and $z$, over the undeformed configuration of the plates, immediately following fracture nucleation, for one of the values of the regularization length, $\varepsilon=0.016$ mm in Fig.~\ref{Fig5}(c) and $\varepsilon=0.08$ mm in Fig.~\ref{Fig6}(c).

The results in Figs.~\ref{Fig5}(a) and \ref{Fig6}(a) make it plain that the \texttt{AT}$_1$ model is \emph{not} a viable candidate to describe --- and hence predict --- fracture in general, as it does not agree with the exact results (\ref{Response-Bi-Lin}) and (\ref{Response-Bi-NH}) for the values ($\varepsilon=0.16$ mm and $\varepsilon=0.21$ mm) of the regularization length $\varepsilon$ for which this model was fitted to describe the preceding uniaxial tension tests. In this regard, we recall here from the works of \cite{KBFLP20} and \cite{KDLP25} that the formulas
\begin{equation*}
\sbs^{\texttt{AT}_1}=\sqrt{\dfrac{3 G_c E}{16(1-\nu)\varepsilon}}\quad {\rm and}\quad \sbs^{\texttt{AT}_1}=\mu\left(\dfrac{l^6-1}{l^5}+\dfrac{2\mu(1-l^4)}{\Lambda l^9}\right),
\end{equation*}
where $l$ is the root closest to $1$ of the non-linear algebraic equation $f_{\texttt{bs}}^{\texttt{AT}_1}(l;\varepsilon)=\mu(2l^2+1/l^4-3-\mu(l^4-1)^2/(3\Lambda l^8))-3G_c/(8\varepsilon)=0$, provide estimates for the critical values of the biaxial stress $S$ at which the \texttt{AT}$_1$ model predicts fracture nucleation in the plates. 

On the other hand, the results in Figs.~\ref{Fig5}(b) and \ref{Fig6}(b) show that the phase-field model predicts accurately the exact results (\ref{Response-Bi-Lin}) and (\ref{Response-Bi-NH}) in their entirety, provided that the value of the regularization length $\varepsilon$ is chosen to be sufficiently small. 

\subsection{Torsion test}

\begin{figure}[t!]
\centering
\centering\includegraphics[width=0.85\linewidth]{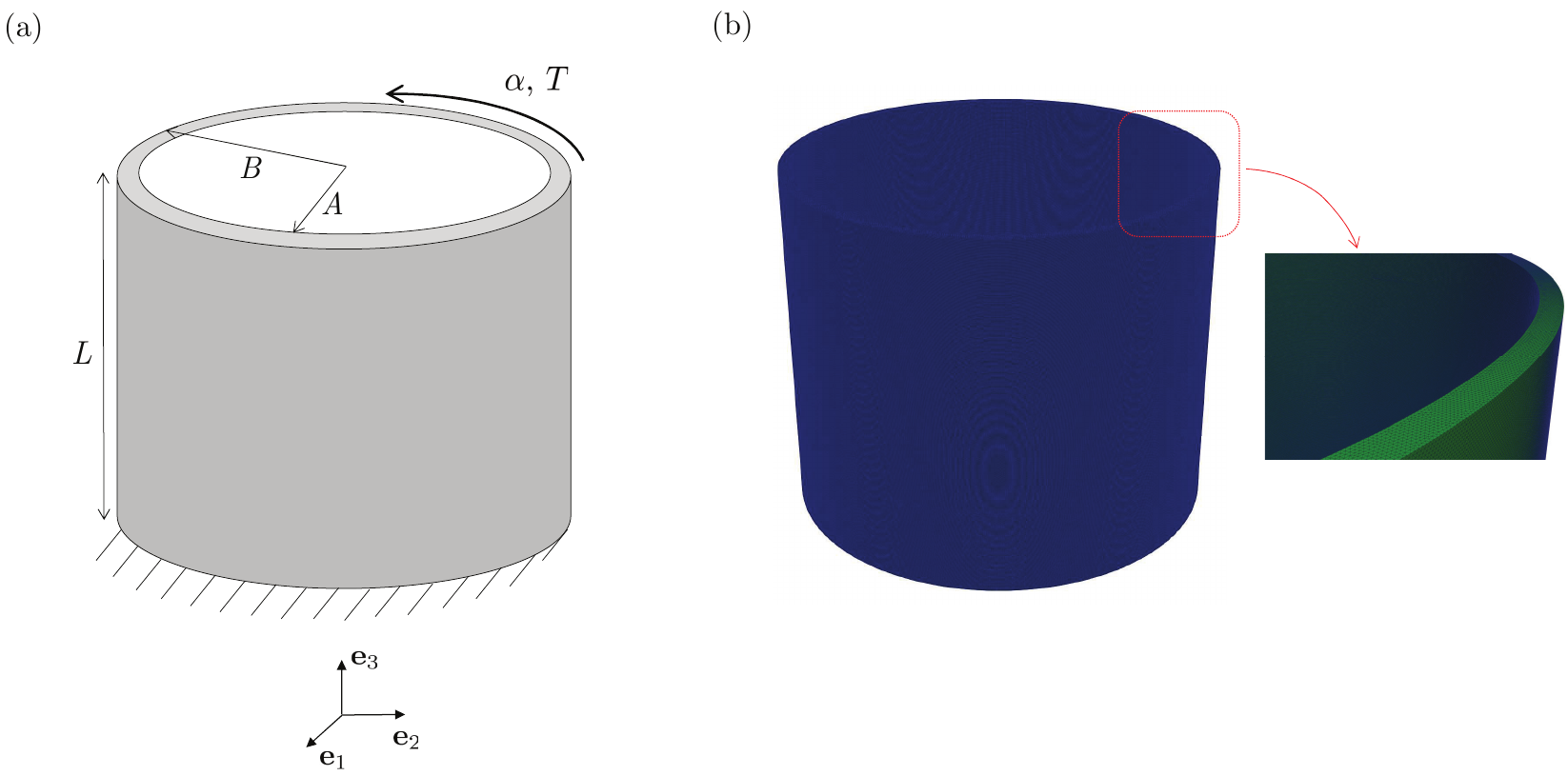}
\caption{\small (a) Schematic of the torsion test. The specimen dimensions are $L=5$ mm, $A=2.85$ mm, and $B=3$ mm. (b) Unstructured FE mesh of uniform element size $\texttt{h}=0.015$ mm utilized for the phase-field simulations of the test.}\label{Fig7}
\end{figure}

The third challenge problem, shown schematically in Fig.~\ref{Fig7}(a), is that of a thin-walled circular tube, of initial length $L=5$ mm, inner radius $A=2.85$ mm, and outer radius $B=3$ mm, that is subjected to torsion by the application of a small angle of twist $\alpha$ at one of its ends, while the opposite end is held fixed. The tube, which can be viewed as the gauge section in a larger specimen, undergoes a state of shear strain and shear stress of the form 
\begin{equation}
\bfE=\dfrac{\alpha X_1}{2L}(\bfe_2\otimes\bfe_3+\bfe_3\otimes\bfe_2)-\dfrac{\alpha X_2}{2L}(\bfe_1\otimes\bfe_3+\bfe_3\otimes\bfe_1)=\gamma(\bfe_{\Theta}\otimes\bfe_Z+\bfe_Z\otimes\bfe_{\Theta})\label{E-gamma}
\end{equation}
and 
\begin{align}
\bfS=\dfrac{2T X_1}{\pi(B^4-A^4)}(\bfe_2\otimes\bfe_3+\bfe_3\otimes\bfe_2)-\dfrac{2TX_2}{\pi(B^4-A^4)}(\bfe_1\otimes\bfe_3+\bfe_3\otimes\bfe_1)\label{S-tau}
=\tau(\bfe_{\Theta}\otimes\bfe_Z+\bfe_Z\otimes\bfe_{\Theta})
\end{align}
for $\bfX\in\Omega=\{\bfX:A^2<X_1^2+X_2^2<B^2,\,0<X_3<L\}$, where $\gamma=\alpha R/(2L)$ is the shear strain, $R=\sqrt{X_1^2+X_2^2}$, $T$ stands for the resultant torque at the ends of the tube, $\tau=2T R/(\pi(B^4-A^4))$ is the shear stress, and where $\{\bfe_1,\bfe_2,\bfe_3\}$ is the Cartesian laboratory frame of reference indicated in the figure, while $\{\bfe_R,\bfe_{\Theta},\bfe_Z\}$ stands for the corresponding cylindrical frame of reference. 

While the strain (\ref{E-gamma}) and stress (\ref{S-tau}) are not exactly spatially uniform, they are nearly so because of the small thickness $t=B-A=0.15$ mm of the tube relative to its outer radius $B=3$ mm. One can thus consider that the tube undergoes an approximately uniform shear stress of average value $S=T (A+B)/(\pi(B^4-A^4))=T(A-t)/(2\pi A^3 t)+O(t)$ and corresponding uniform shear strain of average value $\alpha (A+B)/(4L)=\alpha(2A+t)/(4L)$. This state of roughly uniform stress and strain remains so until the shear stress measure $S$ reaches the shear strength $\sss$ of the material, at which point the tube is severed by the abrupt nucleation of cracks of arbitrary location that are oriented at $45^\circ$ with respect to the symmetry axis of the tube, $\bfe_3$ in Fig.~\ref{Fig7}(a). The arbitrariness in crack location is again due to the inherent stochasticity of strength as a material property. The orientation at $45^\circ$ is the result of the maximum principal stress being aligned in that direction.

For the case when the tube is made of the soda-lime glass, the measures of stress and strain are related according to
\begin{equation}
S=\left\{\hspace{-0.15cm}\begin{array}{ll}
\mu\dfrac{\alpha (A+B)}{2L}& {\rm if}\;\; 0\leq \alpha<\alpha_{\texttt{ss}}\vspace{0.2cm}\\
0&{\rm if} \;\;\alpha_{\texttt{ss}}\leq \alpha\end{array}\right. \label{Response-Shear}
\end{equation}
where $\alpha_{\texttt{ss}}$ denotes the value of the angle of twist at which $S=\sss$. We do not consider the case when the tube is made of the PU elastomer, as not only the Poynting effect but also buckling take place and the stress and strain lose their uniformity in space before fracture occurs. Once more, for a computational model of fracture to be viable, it must be able to deliver predictions that agree with the result (\ref{Response-Shear}), in addition to being able to deliver predictions that agree with the previous results  (\ref{Response-Uni})-(\ref{Response-Bi-NH}) for uniaxial and biaxial tension.

\begin{figure}[t!]
\centering
\centering\includegraphics[width=0.99\linewidth]{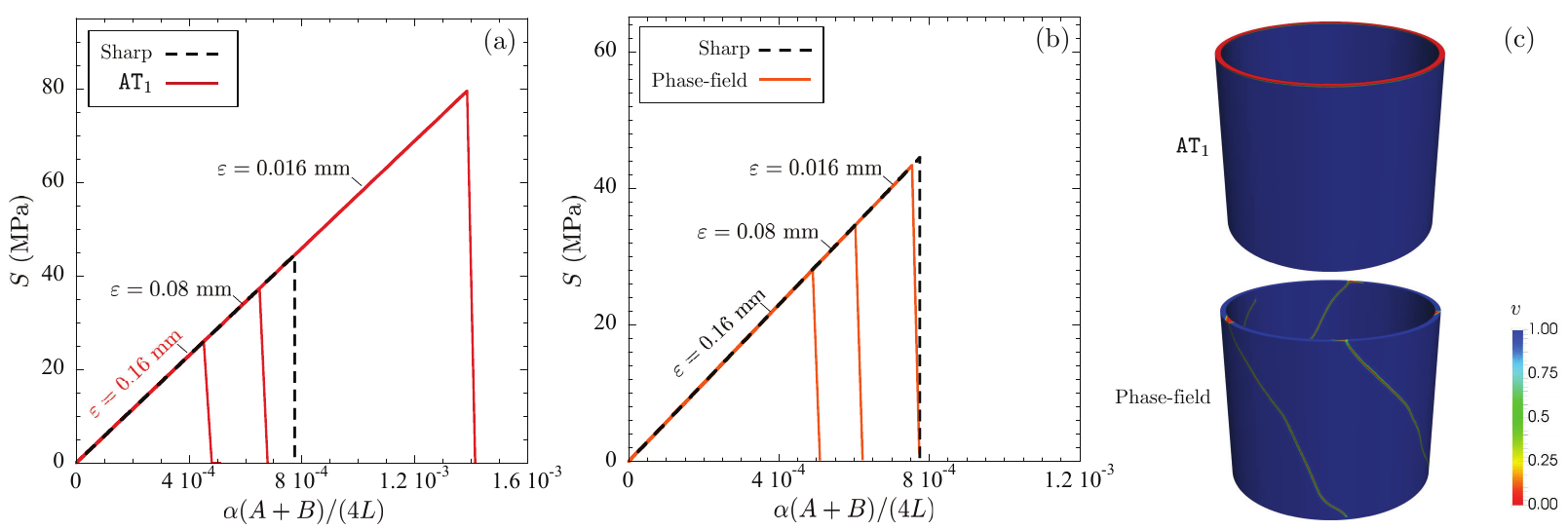}
\caption{\small Torsion test of a soda-lime glass thin-walled circular tube.  Comparisons between the exact result (\ref{Response-Shear}) for the stress-strain response of the tube and the predictions by (a) the \texttt{AT}$_1$ and (b) the phase-field models for three different values of the regularization length $\varepsilon$. (c) Contour plots of the phase field $v$ over the undeformed configuration of the tube right after fracture nucleation, as predicted by the \texttt{AT}$_1$ and phase-field models for $\varepsilon=0.016$ mm.}\label{Fig8}
\end{figure}

Figure \ref{Fig7}(b) displays the FE mesh utilized to carry out the phase-field simulations of the torsion test on the soda-lime glass tubes. The results are presented in Fig.~\ref{Fig8}. In keeping with the format used for the two previous challenge problems governed by strength, Figs.~\ref{Fig8}(a) and \ref{Fig8}(b) compare the predictions generated by the \texttt{AT}$_1$ and phase-field models for the stress-strain response ($S$ vs. $\alpha (A+B)/(4L)$) of the tube with the exact result (\ref{Response-Shear}) for three values of the regularization length $\varepsilon$, while   Fig.~\ref{Fig8}(c) provides contour plots for the phase field $v$ right after nucleation, as predicted by both models for $\varepsilon=0.016$ mm.

Consistent with those in Figs.~\ref{Fig5} and \ref{Fig6}, the results in Fig.~\ref{Fig8} further illustrate the inability of the \texttt{AT}$_1$ model to describe correctly fracture nucleation governed by material strength. Indeed, the critical shear stress, say $\sss^{\texttt{AT}_{1}}$, at which the \texttt{AT}$_1$ model predicts fracture nucleation in the tube is approximately given by the formula \citep{KBFLP20}
\begin{equation*}
\sss^{\texttt{AT}_1}=\sqrt{\dfrac{3 G_c E}{16(1+\nu)\varepsilon}}.
\end{equation*}
When evaluated at the regularization length $\varepsilon=0.16$ mm for which this model was fitted to describe the preceding uniaxial tension test for soda-lime glass, the value for $\sss^{\texttt{AT}_1}$ disagrees significantly ($\sss^{\texttt{AT}_1}=0.58\, \sss$) with the actual shear strength $\sss$ of the material. What is more, the model predicts the nucleation of a crack that is perpendicular to the applied angle of twist, as opposed to the expected $45^\circ$ orientation.

By contrast, consistent with those in Figs.~\ref{Fig2}, \ref{Fig3}, \ref{Fig5}, and \ref{Fig6}, the results in Fig.~\ref{Fig8} show once again that the phase-field model predicts accurately where and when fracture nucleates in the tube, provided that the value of the regularization length $\varepsilon$ is chosen to be sufficiently small. 

\section{Fracture nucleation governed by Griffith energy competition}\label{Sec: Griffith Nucleation}

After decades of investigations centered around fracture nucleation in uniaxial tensile tests, on specimens without pre-existing cracks, the combined experimental and theoretical work of \cite{Griffith21} steered the focus of the field of fracture mechanics into a very different place, that of the growth of large pre-existing cracks. This involved both the study of the critical loads at which the front of large pre-existing cracks start to grow, as well as the subsequent propagation of such cracks. In this section, we present a challenge problem that deals with the former, in other words, we present a challenge problem that probes fracture nucleation from a large pre-existing crack. 

Since the pioneering experiments of \cite{Griffith21} on glass and those of \cite{RT53} on natural rubber, a multitude of experiments have been conducted on a variety of other ceramics and elastomers, as well as on numerous polymers and metals. The consensus is that the Griffith criticality condition
\begin{equation}
-\dfrac{\partial\mathcal{P}}{\partial\Gamma}=G_c\label{Griffith-Cond}
\end{equation}
is a necessary condition for the nucleation of fracture from the front of large pre-existing cracks in nominally isotropic elastic brittle materials subjected to quasi-static mechanical loads. In this condition, the left-hand side $-\partial \mathcal{P}/\partial \mathrm{\Gamma}$ denotes the change in potential energy (i.e., the total stored elastic energy minus the work done by the external forces) in the specimen at hand with respect to an added surface area ${\rm d}\mathrm{\Gamma}$ to an existing crack in its reference state, while the right-hand side $G_c$ stands for the critical energy release rate or toughness, a macroscopic material property. Physically, the Griffith criticality condition is one of energy competition between bulk deformation energy and surface fracture energy: under a given load and for a preset crack path, a crack will have a particular surface area $\Gamma$ in its reference state if any putative added surface would result in an expenditure of surface energy (assumed to be proportional to the added surface) greater than the accompanying decrease in potential energy $\mathcal{P}$.

Over the years, significant efforts have been devoted to design standardizable experiments that allow for a straightforward calculation of the derivative $-\partial \mathcal{P}/\partial \mathrm{\Gamma}$ so that, in conjunction with the Griffith criticality condition (\ref{Griffith-Cond}), one could measure the critical energy release rate $G_c$ for a given material of interest. A list of such experiments focused on hard materials can be found, for instance, in the classical handbook by \cite{Tada73}. For soft materials, the experiments originally proposed by \cite{RT53} remain the more widely used. Among all of these experiments, the so-called ``pure-shear'' fracture test is arguably the simplest to carry out and analyze, as well as the most versatile in that it can be used for both hard and soft materials the same \citep{RT53,Knauss1966,Rice1968}. Consequently, we choose this test as the challenge problem that any viable computational model of fracture must convincingly handle in its characterization of fracture nucleation from a large pre-existing crack. In the sequel, we spell out the details of this test.

\subsection{Pure-shear fracture test}

\begin{figure}[H]
\centering
\centering\includegraphics[width=0.95\linewidth]{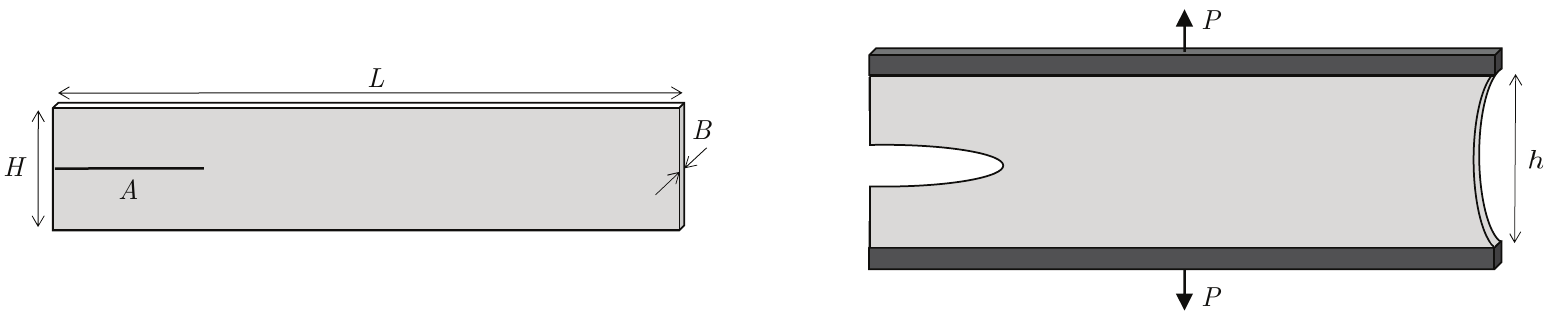}
\caption{\small Schematic of the pure-shear fracture test. The specimen dimensions are $L=50$ mm, $H=5$ mm, $B=0.5$ mm, and $A=10$ mm.}\label{Fig9}
\end{figure}

Figure \ref{Fig9} depicts the pure-shear fracture test chosen here as the fourth challenge problem. It involves a specimen in the form of a thin strip of length $L=50$ mm, height $H=5$ mm, and thickness $B=0.5$ mm that contains a pre-existing edge crack of initial length $A=10$ mm; in other words, the specimen is essentially an infinitely long strip that contains a semi-infinitely long edge crack. The specimen is firmly clamped on its top and bottom and subjected to a prescribed separation $h$ between the grips. The resultant force at the grips is denoted by $P$. 

Thanks to the special specimen geometry and applied boundary conditions, as first worked out by \cite{RT53} for isotropic non-linear elastic materials,\footnote{To be precise, \cite{RT53} worked out the result for isotropic \emph{incompressible} non-linear elastic materials. Their approach remains valid, \emph{mutatis mutandis}, for compressible materials.} say with elastic energy density $\Psi(\lambda_1,\lambda_2,\lambda_3)$ in terms of the principal stretches $\lambda_1,\lambda_2,\lambda_3$, the change in potential energy $-\partial \mathcal{P}/\partial \mathrm{\Gamma}$ in the Griffith criticality condition (\ref{Griffith-Cond}) for this test can be estimated by the simple formula
\begin{equation*}
-\dfrac{\partial\mathcal{P}}{\partial\Gamma}=H\Psi(\lambda,\lambda_l,1)\quad {\rm with}\quad \lambda=\dfrac{h}{H},
\end{equation*}
where $\lambda_l$ is defined implicitly as the root closest to 1 of the non-linear algebraic equation
\begin{equation*}
\dfrac{\partial\Psi}{\partial\lambda_2}(\lambda,\lambda_l,1)=0.
\end{equation*}
Thus, in view of this result, according to (\ref{Griffith-Cond}), there is a critical value $h_{cr}$ of the separation $h$ between the grips at which the crack will start to grow; note that such a critical value is independent of $L$, $B$, and $A$. In particular, the crack will grow straight ahead preserving self-similarity with respect to its initial geometry. 

When specialized to soda-lime glass, the above result indicates that the crack will start to grow at the critical grip separation
\begin{equation}
h_{cr}=\left(1+\sqrt{\dfrac{2(1-\nu^2)G_c}{HE}}\right)H.\label{hcr-Lin}
\end{equation}
For the case of the PU elastomer, crack growth will ensue at
\begin{equation}
h_{cr}=\left(\sqrt{1+\dfrac{G_c}{H\mu}+\dfrac{\sqrt{G_c(G_c+2H\mu)}}{H\mu}}+
\dfrac{\mu}{2\sqrt{\left(\dfrac{1}{H}+\dfrac{2\mu}{G_c}\right)\left(H+\dfrac{G_c}{\mu}+\dfrac{\sqrt{G_c(G_c+2H\mu)}}{\mu}\right)}\Lambda}\right)H+O\left(\Lambda^{-2}\right).\label{hcr-NH}
\end{equation}
Again, viability for a computational model of fracture requires its predictions to agree with these results, in addition to the previous results (\ref{Response-Uni}), (\ref{Response-Bi-Lin}), (\ref{Response-Bi-NH}), and (\ref{Response-Shear}) for uniaxial tension, biaxial tension, and torsion.

\begin{figure}[b!]
\centering
\centering\includegraphics[width=0.9\linewidth]{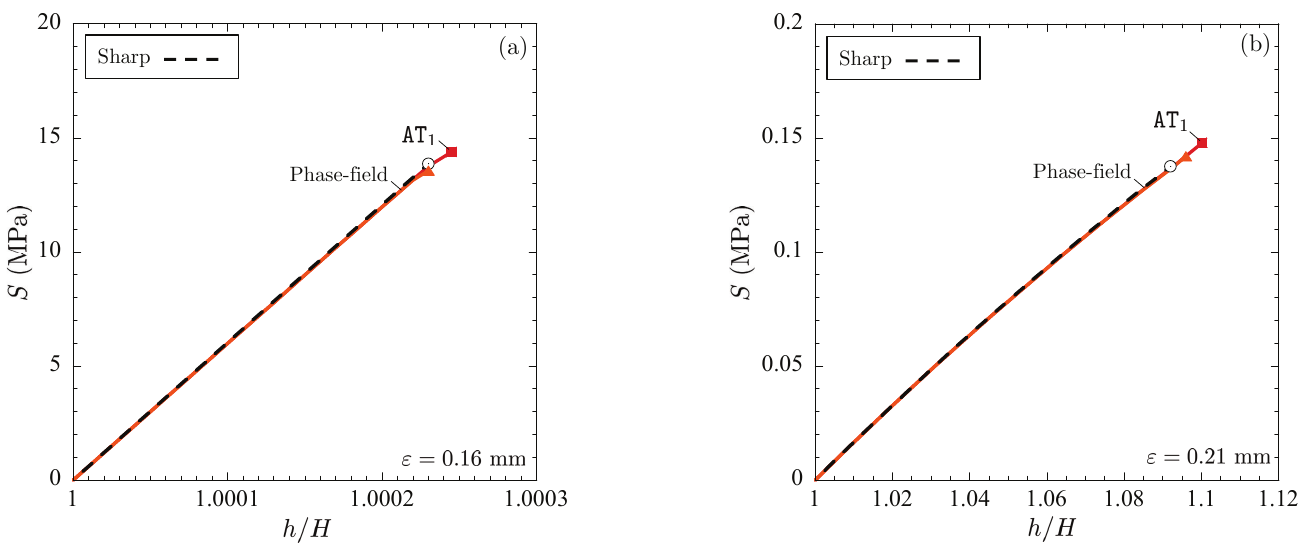}
\caption{\small Comparisons between the exact result for the global stress-stretch response of the specimens and the predictions by the \texttt{AT}$_1$ and phase-field models in pure-shear fracture tests on (a) the soda-lime glass and (b) the PU elastomer. The circles mark the critical separation (\ref{hcr-Lin}) and (\ref{hcr-NH}) between the grips at which the crack starts to grow. The squares and triangles indicate the corresponding predictions by the \texttt{AT}$_1$ and phase-field models, respectively.}\label{Fig10}
\end{figure}
In the simulations of the pure-shear fracture test by means of the \texttt{AT}$_1$ and phase-field models, to reduce computational cost, we exploit symmetry and perform them over a half of the strip. The FE mesh used for these simulations is unstructured and refined around the crack front, where the growth of the crack is expected to occur, with elements of size $\texttt{h}=0.05$ mm there. Figures \ref{Fig10}(a) and \ref{Fig10}(b) compare the global stress-stretch response ($S=P/(L B)$ vs. $h/H$) of the specimens as predicted by the \texttt{AT}$_1$ and phase-field models with the corresponding exact results for the soda-lime glass and PU elastomer, respectively. The predictions pertain to the values $\varepsilon=0.16$ mm and $\varepsilon=0.21$ mm for the regularization length. Smaller values of $\varepsilon$ were checked to yield essentially the same results. The critical grip separations signaling the onset of crack growth are marked directly on the plots, by squares for the \texttt{AT}$_1$ model and by triangles for the phase-field model. For direct comparison, the exact results (\ref{hcr-Lin}) and (\ref{hcr-NH}) are marked by circles.

The main observation from Fig.~\ref{Fig10} is that the phase-field model again predicts accurately the exact results (\ref{hcr-Lin}) and (\ref{hcr-NH}). 

Incidentally, while the two preceding challenge problems have already made clear that the \texttt{AT}$_1$ model is not a viable model, Fig.~\ref{Fig10} shows that this model delivers predictions that are in good agreement (within a few percent error) with the exact results (\ref{hcr-Lin}) and (\ref{hcr-NH}). This should not come as a surprise since the \texttt{AT}$_1$ model was originally conceived as a regularization of the variational theory of sharp fracture of \cite{Francfort98}, which aims at providing a mathematical statement for Griffith energy competition.

\section{Fracture nucleation governed by the mediation between strength and Griffith}\label{Sec: Mediation Nucleation}

The four challenge problems discussed above have dealt with nucleation of fracture in two opposite limiting cases: when the stress field is roughly spatially uniform and when the stress field is strongly non-uniform (in fact, singular) because of the presence of a large crack. In this section, we turn to challenge problems that deal with fracture nucleation under the remaining situations between these two opposite limiting conditions, when the stress field is not spatially uniform, but not as non-uniform as around large pre-existing cracks. These include fracture nucleation from notches, smooth and sharp, small pre-existing cracks, and from any other subregion in a structure under a non-uniform state of stress.

Since the 1930s, numerous experiments on specimens, made of ceramics, elastomers, polymers, and metals alike, featuring U and V notches \citep{Shand54,Petch54,Greensmith60,Andrews63,Dunn97,Gomez05}, as well as specimens featuring small pre-existing edge cracks \citep{Busse34,RT53,Thomas1970,Kimoto85,Ritchie04,Chen17} have repeatedly shown that nucleation of fracture from the front of the notch or crack is the result of a mediation between material strength and Griffith energy competition. The same is true for fracture nucleation that occurs in any other subregion --- entirely within the bulk or including a part of the boundary --- where the stress is non-uniform, such as in indentation tests \citep{Roesler56,Mouginot85,Lawn98} and Brazilian tests \citep{Sato79,Bisai19,Sheikh19} in hard materials and in poker-chip tests \citep{Busse38,GL59,Creton10,Euchler20,KLP21,GuoRavi23} and related tests \citep{GentPark84,Poulain17,Poulain18,BCLLP24} in soft materials. 

Although there is currently no sharp fracture theory that can describe in a complete and \emph{quantitative} manner the mediation between material strength and Griffith energy competition that has been observed experimentally, there are plenty of well-established \emph{qualitative} results. Any viable computational model of fracture must be able to reproduce these results. Among the various experiments that have been developed to probe such qualitative results, we choose here as the next three challenge problems: the single edge notch test, the indentation test of a block with a flat-ended cylindrical punch, and the poker-chip test. In part, these tests are selected for their robustness and widespread use; see, e.g., \cite{Greensmith63,Kimoto85,Mouginot85}, and \cite{Creton10}. More critically, they are selected because they span different subsets of the material strength surface, in particular, subsets around: uniaxial tensile strength, shear strength, and hydrostatic strength. The next three subsections provide the pertinent details for each of these challenge problems along with their analyses.

\subsection{Single edge notch test}

\begin{figure}[t!]
\centering
\centering\includegraphics[width=0.90\linewidth]{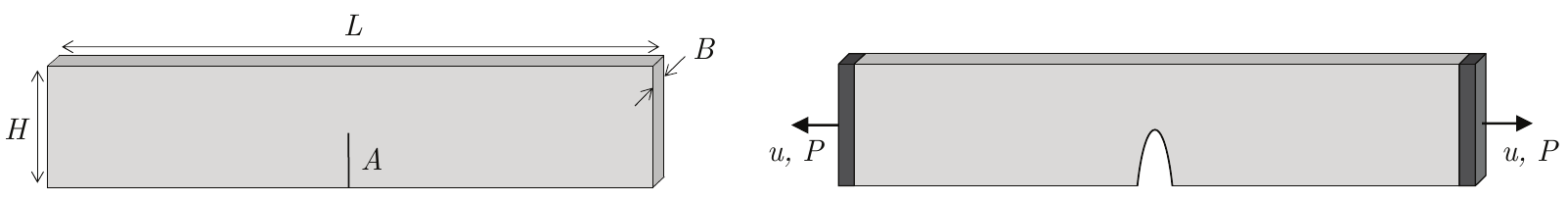}
\caption{\small Schematic of the single edge notch test. The specimens' dimensions are $L=25$ mm, $H=5$ mm, $B=0.25$ mm, and $A=0.025,0.1,0.5,1,1.5$ mm.}\label{Fig11}
\end{figure}

The fifth challenge problem is the single edge notch test depicted in Fig.~\ref{Fig11}. It consists of strips of length $L=25$ mm, width $H=5$ mm, and thickness $B=0.25$ mm that contain a pre-existing edge crack of lengths $A=0.025, 0.1, 0.5, 1,$ and $1.5$ mm. The strips are firmly clamped on their left and right boundaries and these pulled apart by an applied displacement $u$. The resultant force is denoted by $P$ and the corresponding global stress by $S=P/(BH)$. As $u$ is increased, the strips deform elastically until a critical value $u_{cr}$ is reached at which point the crack starts to grow straight ahead. The corresponding critical value $S_{cr}$ of the global stress $S$ at this fracture nucleation event is bounded according to
\begin{equation}
S_{cr}\leq\min\left\{\sts,S_{G}\right\},\label{Scr-SENT}
\end{equation}
where $S_{G}$ is the value of $S$ at which the Griffith criticality condition (\ref{Griffith-Cond}) is satisfied. In particular, the equality $S_{cr}=S_{G}$ is attained for sufficiently large cracks, while $S_{cr}=\sts$ holds for sufficiently small cracks. For cracks of intermediate length, the inequality in (\ref{Scr-SENT}) is strict.

\begin{figure}[b!]
\centering
\centering\includegraphics[width=0.9\linewidth]{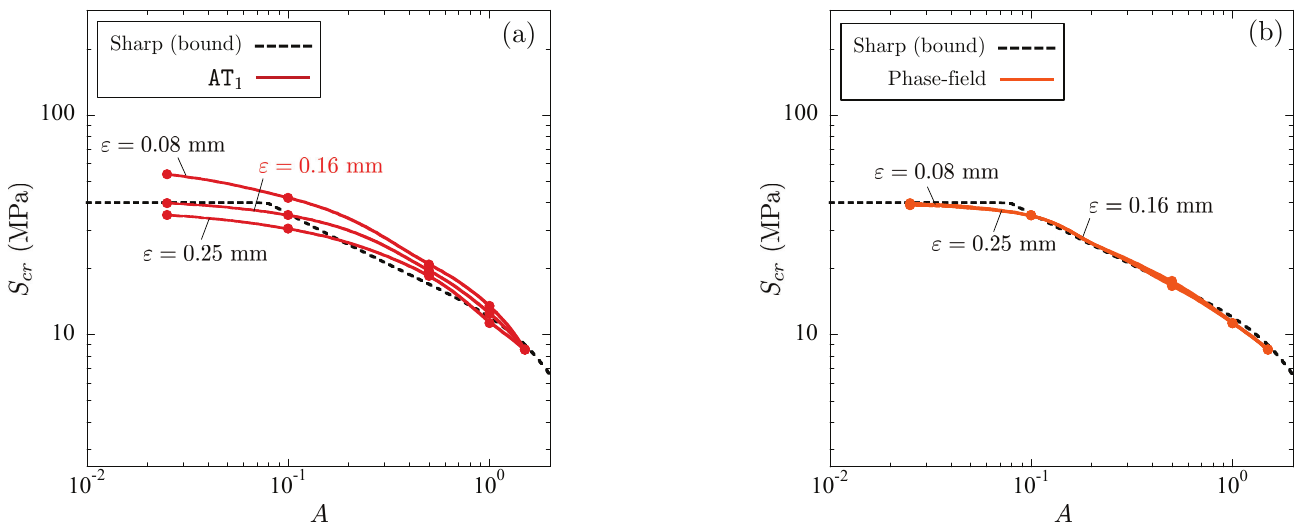}
\caption{\small Single edge notch test on the soda-lime glass. Comparisons between the bound (\ref{Scr-SENT-glass}) for the critical stress $S_{cr}$ at fracture nucleation and the predictions by (a) the \texttt{AT}$_1$ and (b) phase-field models for three values of the regularization length $\varepsilon$. The results are shown as a function of the crack length $A$.}\label{Fig12}
\end{figure}
\begin{figure}[t!]
\centering
\centering\includegraphics[width=0.9\linewidth]{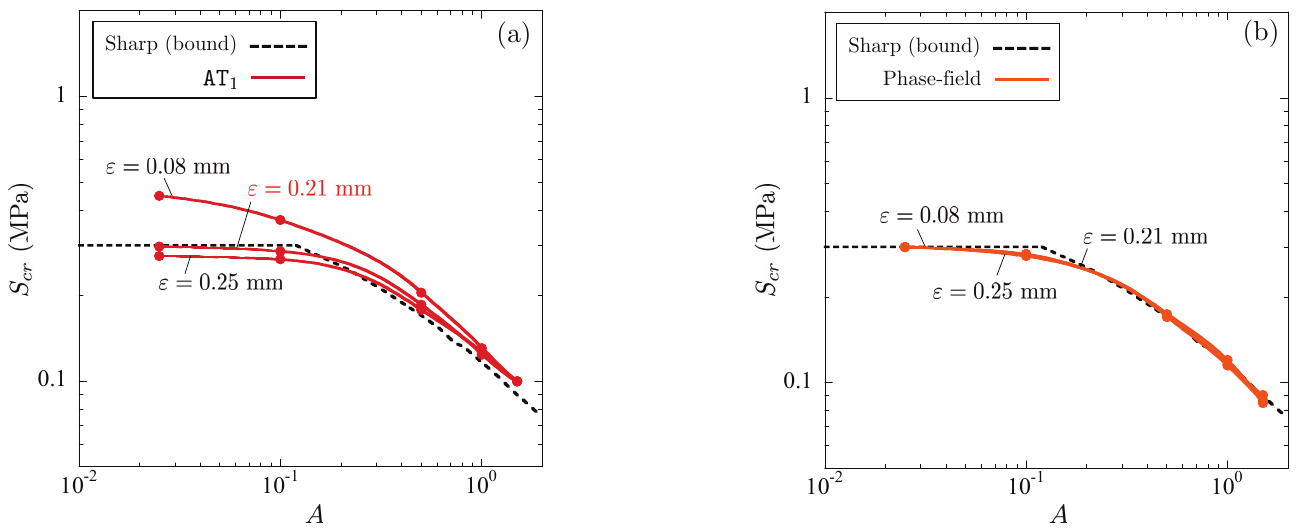}
\caption{\small Single edge notch test on the PU elastomer. Comparisons between the bound for the critical stress $S_{cr}$ at fracture nucleation computed from full-field FE simulations and the predictions by (a) the \texttt{AT}$_1$ and (b) phase-field models for three values of the regularization length $\varepsilon$. The results are shown as a function of the crack length $A$.}\label{Fig13}
\end{figure}

When specialized to soda-lime glass, a modification of a classical estimate for $S_{G}$ \citep{Tada73} that accounts for the applied grip boundary conditions considered here allows to write the bound (\ref{Scr-SENT}) in the closed form

\begin{equation}
S_{cr}\leq\min\left\{\sts,\dfrac{\cos\left(\dfrac{\pi A}{2H}\right)\sqrt{\dfrac{G_c E}{\pi A}}}{\left(0.752+1.0431\dfrac{A}{H}+0.6076\left(1-\sin\left(\dfrac{\pi A}{2H}\right)\right)^3\right)\sqrt{\dfrac{2H}{\pi A}\tan\left(\dfrac{\pi A}{2H}\right)}}\right\}.\label{Scr-SENT-glass}
\end{equation}
Although no analogous estimate exists for non-linear elastic brittle materials, standard FE calculations can be readily used to numerically determine the corresponding results for $S_G$ and hence the corresponding bound for $S_{cr}$. We carry out such FE calculations in our analysis of the single edge notch test for the representative PU elastomer under study in this work.

To reduce computational cost, we exploit symmetry and carry out the simulations of the single edge notch test by means of the \texttt{AT}$_1$ and phase-field models over a half of the strips. The FE meshes are unstructured and refined around the crack front with smallest element size $\texttt{h}=0.005$ mm. Figure \ref{Fig12} presents comparisons between the bound (\ref{Scr-SENT-glass}) and the results for $S_{cr}$ predicted by the \texttt{AT}$_1$ and phase-field models for soda-lime glass. The results are shown as a function of the crack length $A$ and pertain to three different values of the regularization length $\varepsilon$. Figure \ref{Fig13} presents analogous comparisons for the PU elastomer. In those, as noted above, the bound for $S_{cr}$ is computed from full-field FE simulations.

A quick glance at Figs.~\ref{Fig12} and \ref{Fig13} suffices to recognize that the phase-field model delivers predictions that not only satisfy the bound for $S_{cr}$, but in particular they are such that $S_{cr}\nearrow S_{G}$ for increasing $A$ and $S_{cr}\nearrow\sts$ for decreasing $A$, consistent with the exact result, this irrespective of the value of the regularization length $\varepsilon$. 

The plots also show that the \texttt{AT}$_1$ model happens to generate results that are in compliance (to within a few percent error) with the bound for $S_{cr}$ in the limits when $A$ is small and large when $\varepsilon=0.16$ mm  for the soda-lime glass and  $\varepsilon=0.21$ mm for the PU elastomer, that is, when the value for the regularization length is fitted to the correct uniaxial tensile strength $\sts$.

\subsection{Indentation test of a block with a flat-ended cylindrical punch}

\begin{figure}[t!]
\centering
\centering\includegraphics[width=0.85\linewidth]{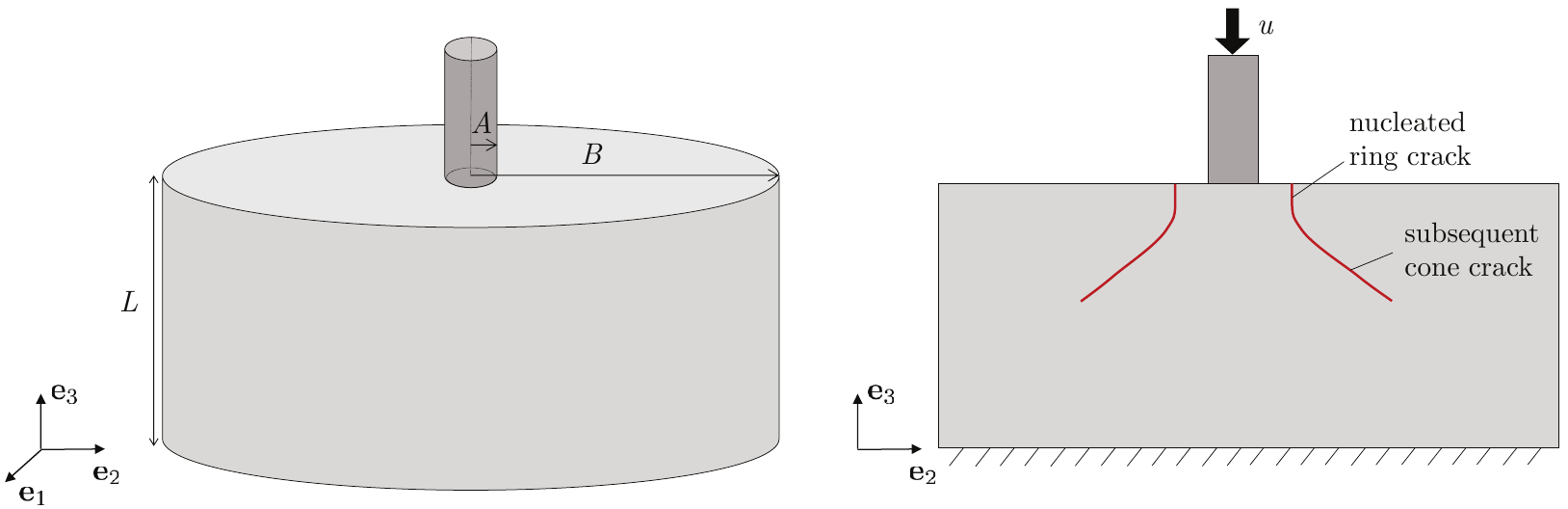}
\caption{\small Schematic of the indentation test. The specimen dimensions are $L=25$ mm and $B=25$ mm. The radius of the indenter is $A=1$ mm.}\label{Fig14}
\end{figure}

The sixth challenge problem, shown schematically in Fig.~\ref{Fig14}, is that of a cylindrical block, of length $L=25$ mm and radius $B=25$ mm, that is indented on its top boundary along its symmetry axis by a flat-ended cylindrical punch of radius $A=1$ mm, while its bottom boundary rests on a rigid substrate. The punch is mechanically rigid and it indents the block by moving with a prescribed displacement $u$. 

\begin{figure}[t!]
\centering
\centering\includegraphics[width=0.99\linewidth]{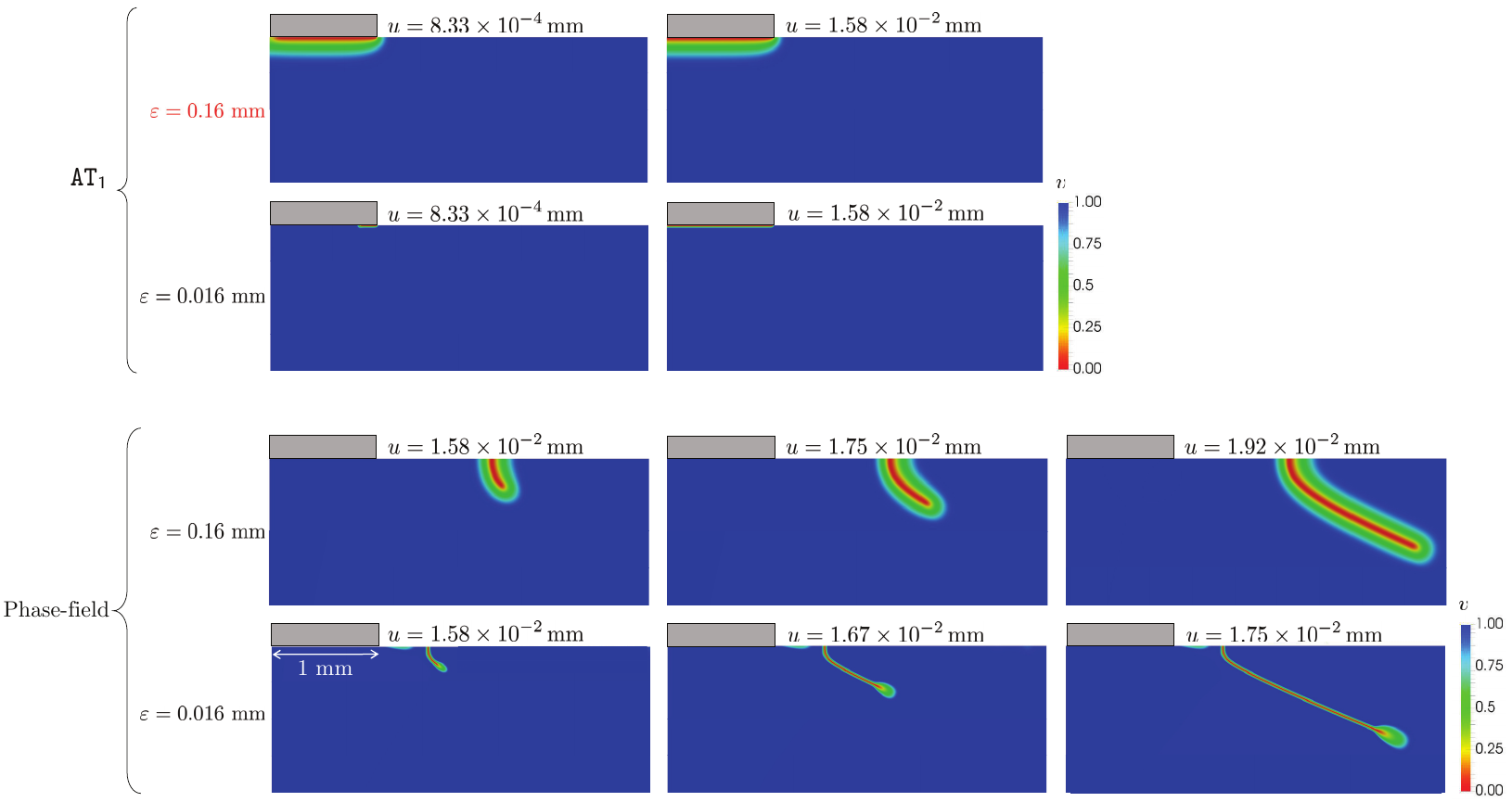}
\caption{\small Indentation test on the soda-lime glass. Contour plots of the phase field $v$ over the undeformed configuration of the specimen at two and three applied displacements $u$, as predicted by the \texttt{AT}$_1$ and phase-field models for two values of the regularization length $\varepsilon$.}\label{Fig15}
\end{figure}

For a linear elastic brittle material, as $u$ is increased, the block deforms elastically until a critical value $u_{cr}$ is reached at which point a crack in the form of a ring is nucleated from the surface of the specimen near the indenter. The radius of this ring crack is larger than the radius $A$ of the indenter, while its depth is smaller. As $u$ is increased further, the nucleated ring crack turns and grows at a roughly constant angle with respect to the surface of the specimen, thus forming a cone crack. Figure \ref{Fig14} schematically shows these nucleation and subsequent propagation of fracture events. For a computational model of fracture to be viable, it must be able to deliver predictions that agree with these \emph{qualitative} results. 

We remark that for the case when the block is made of a soft non-linear elastic brittle material, the fracture process is very different to that described above for a linear elastic brittle material. It is a process of puncturing with friction. This is beyond the scope of this work and hence we do not consider it here.

To simulate the indentation test on soda-lime glass using the \texttt{AT}$_1$ and phase-field models, we fully exploit symmetry and formulate the test as a 2D axisymmetric problem. The FE mesh used to perform the simulations is, as always, unstructured and refined in the region around the indenter, with smallest element size $\texttt{h}=0.005$ mm, where the crack is expected to nucleate and propagate. Figure \ref{Fig15} presents contour plots for the phase field $v$ over the undeformed configuration of the specimen as predicted by the \texttt{AT}$_1$ and phase-field models for regularization lengths $\varepsilon=0.16$ mm and $0.016$ mm. The plots correspond to two and three different values of the applied indentation displacement $u$, the first of these right after fracture nucleation, and the others at larger displacements.

It is immediately apparent from Fig.~\ref{Fig15} that the phase-field model predicts the correct nucleation of a ring crack and its subsequent growth into a cone crack upon further loading. This is not the case for the  \texttt{AT}$_1$ model, which does not even predict the nucleation of a crack, but instead predicts a non-physical diffused region of damage underneath the indentor. 

The results in Fig.~\ref{Fig15} also serve to illustrate that for any regularized computational model of fracture, a sufficiently small regularization length relative to the radius $A$ of the indenter is required to properly handle this challenge problem, since the thickness of the regularized crack needs to be much smaller than $A$. Notably, the regularization length $\varepsilon=0.16$ mm obtained by fitting the \texttt{AT}$_1$ model to the uniaxial tensile strength $\sts=40$ MPa of soda-lime glass is not quite small enough.

\subsection{Poker-chip test}

\begin{figure}[H]
\centering
\centering\includegraphics[width=0.90\linewidth]{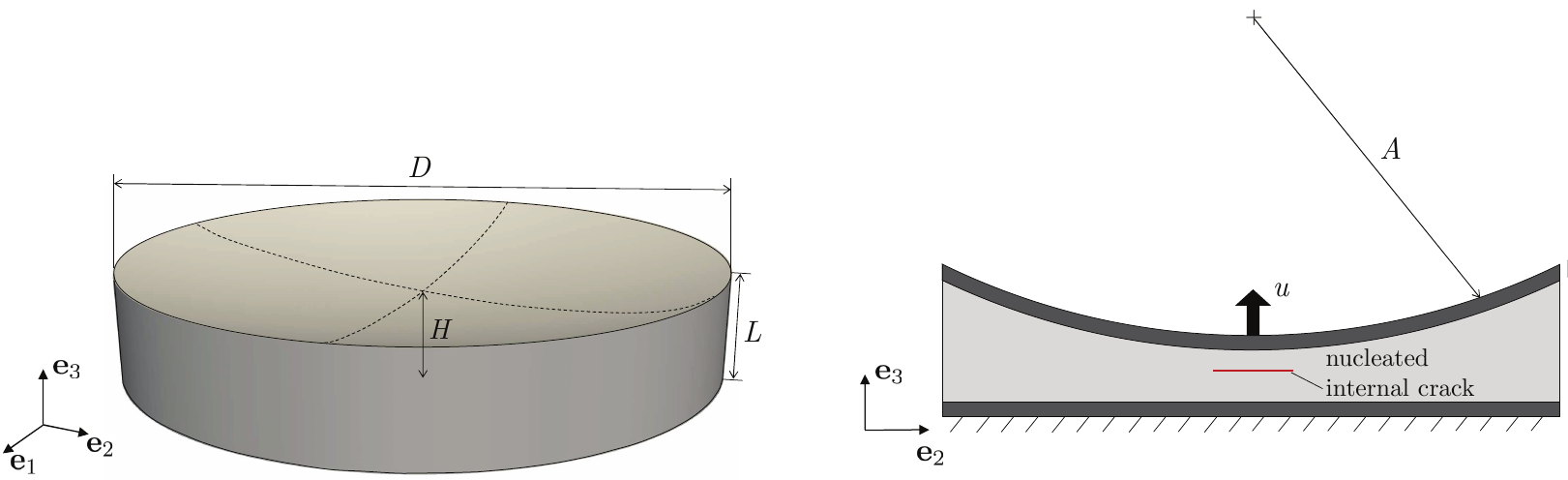}
\caption{\small Schematic of the poker-chip test. The specimen consists of a circular disk of initial diameter $D=10$ mm and radially increasing thickness from $H=1$ mm to $L=1.7$ mm that is firmly bonded between a flat substrate and a spherical fixture of radius $A=18.2$ mm.}\label{Fig16}
\end{figure}

The seventh challenge problem is the poker-chip test shown schematically in Fig.~\ref{Fig16}. The specimen is made of a circular disk of diameter $D=10$ mm, whose bottom boundary is fixed to a flat substrate and whose top boundary is firmly bonded to a spherical fixture\footnote{The finite curvature of the top fixture is a modification of the typical flat geometry found in classical poker-chip tests \citep{Busse38,GL59,Euchler20,GuoRavi23}. It allows to keep the hydrostatic stress that leads to fracture nucleation more localized around the centerline of the disk.} of radius $A=18.2$ mm. The thickness of the disk is hence not constant but increases radially from $H=1$ mm to $L=1.7$ mm; this particular geometry was used by \cite{Creton10} in their poker-chip experiments on various PU elastomers. The spherical fixture is moved vertically by a prescribed displacement $u$. 

For a nearly incompressible elastic brittle material, the applied displacement $u$ leads to a non-homogeneous state of stress within the disk that features a large hydrostatic component around its axis of symmetry or centerline. It is under this state of stress that a crack perpendicular to the applied displacement nucleates within the disk around its centerline at some critical value $u_{cr}$ of $u$. For a computational model of fracture to be viable, it must be able to deliver predictions that agree with these \emph{qualitative} results. 

For the case when the disk is made of a highly compressible elastic brittle material, the fracture process may be very different to that described above for a nearly incompressible elastic brittle material. In fact, as far as we know, essentially all data from poker-chip tests available in the literature pertains to nearly incompressible soft materials. For this reason, we only consider the poker-chip test on the PU elastomer and not on the soda-lime glass.

\begin{figure}[t!]
\centering
\centering\includegraphics[width=0.95\linewidth]{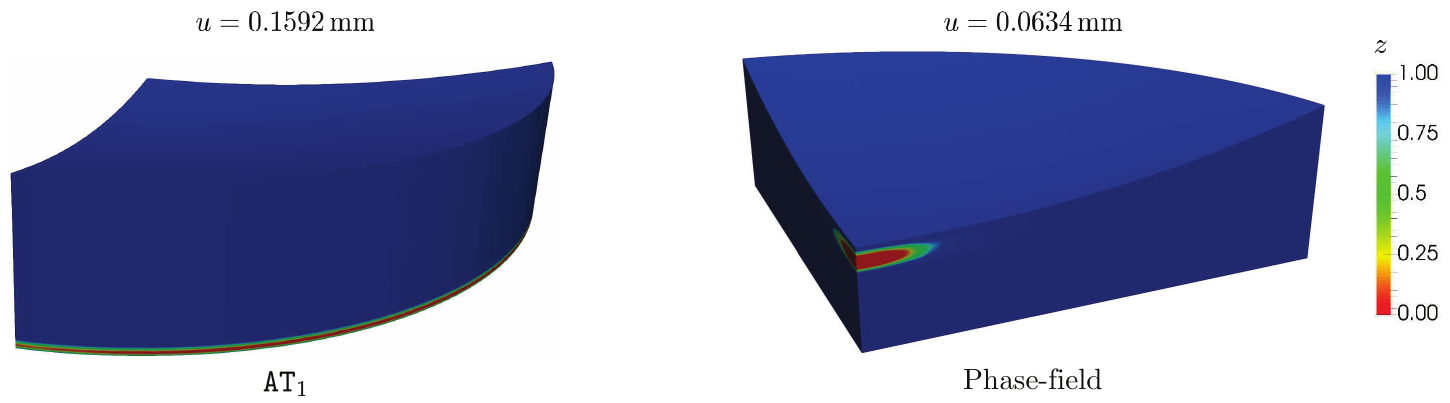}
\caption{\small Poker-chip test on the PU elastomer. Contour plots of the phase field $z$ over the undeformed configuration of the specimen right after fracture nucleation, as predicted by the \texttt{AT}$_1$ and phase-field models for regularization length $\varepsilon=0.025$ mm.}\label{Fig17}
\end{figure}

To simulate the poker-chip test on the PU elastomer by means of the \texttt{AT}$_1$ and phase-field models, we exploit symmetry and carry out the calculations over a quarter of the disk. An unstructured FE mesh, with smallest element size $\texttt{h}=0.005$ mm, that is refined only around the centerline and also around the top and bottom boundaries of the disk suffices to generate converged solutions. Figure \ref{Fig17} presents contour plots for the phase field $z$ over the undeformed configuration of the specimen as predicted by the \texttt{AT}$_1$ and phase-field models right after fracture nucleation. The results correspond to the value $\varepsilon=0.025$ mm for the regularization length, which is small enough relative to the smallest thickness $H=1$ mm of the specimen. Here too, similar to the indentation test, we note that the value $\varepsilon=0.21$ mm for which the \texttt{AT}$_1$ model was fitted to describe the uniaxial tension test is too large relative to $H=1$ mm to represent a regularized crack.

The results in Fig.~\ref{Fig17} show that, yet again, the phase-field model predicts the correct fracture nucleation event. The same is not true about the \texttt{AT}$_1$ model, which predicts the incorrect nucleation of an external crack along the corner of the specimen with the flat substrate, where the elastic energy density (in contrast to the hydrostatic stress) is largest.

\section{Fracture propagation}\label{Sec: Griffith Propagation}

The seven challenge problems discussed above have dealt with nucleation of fracture. In this section, we turn to challenge problems that deal with propagation of fracture. 

While camera systems that track a growing crack's front in time and space could theoretically be used with a multitude of experimental setups to study fracture propagation, in practice, only two classes of experiments have been consistently reliable in this endeavor. One class is the double cantilever beam test \citep{Gilman1960,Gilman64,Moses1968,Ripling67}, including both untapered and tapered variations. These tests allow for the study of Mode I fracture propagation, or crack growth in a purely opening mode. The other class is the trouser test \citep{RT53,Greensmith55}, which allows for the study of Mode III fracture propagation, or crack growth in a purely tearing mode.\footnote{In this work, following \cite{KLP25}, we employ the definition that Mode I, Mode II, and Mode III refer, respectively, to the opening, sliding, and tearing modes of crack growth, wherein either the normal stresses, or the in-plane shear stresses, or the out-of-plane shear stresses \emph{dominate} the stress field around the crack front. This definition generalizes to arbitrary material behavior the classical definition of Mode I, Mode II, and Mode III associated with the asymptotic solution of the equations of \emph{linear} elasto-statics around crack fronts; see, e.g., Chapter 2 in the monograph by \cite{Zehnder2012}.} What sets these experiments apart is that they allow a crack to propagate quasi-statically in a self-similar manner that is directly controlled by the applied boundary conditions, at the same time that they allow for a straightforward analysis of the results. These, but also other more elaborate experiments (see, e.g., \cite{Roesler56}), on a wide range of ceramics, elastomers, polymers, and metals have repeatedly shown that the Griffith criticality condition (\ref{Griffith-Cond}), which we repeat here for symmetry of presentation
\begin{equation}
-\dfrac{\partial\mathcal{P}}{\partial\Gamma}=G_c,\label{Griffith-Cond-Propa}
\end{equation}
is a necessary condition for the propagation of fracture in nominally isotropic elastic brittle materials subjected to quasi-static mechanical loads.

In view of their reliability, popular use, and the crucial fact that they probe two different (opening and tearing) modes of crack growth, we choose a double cantilever beam test and a trousers test as the challenge problems that any viable computational model of fracture must convincingly handle in its characterization of fracture propagation. We detail the specifics of each test and present their analyses in the following two subsections.

\subsection{The double cantilever beam test}

\begin{figure}[H]
\centering
\centering\includegraphics[width=0.90\linewidth]{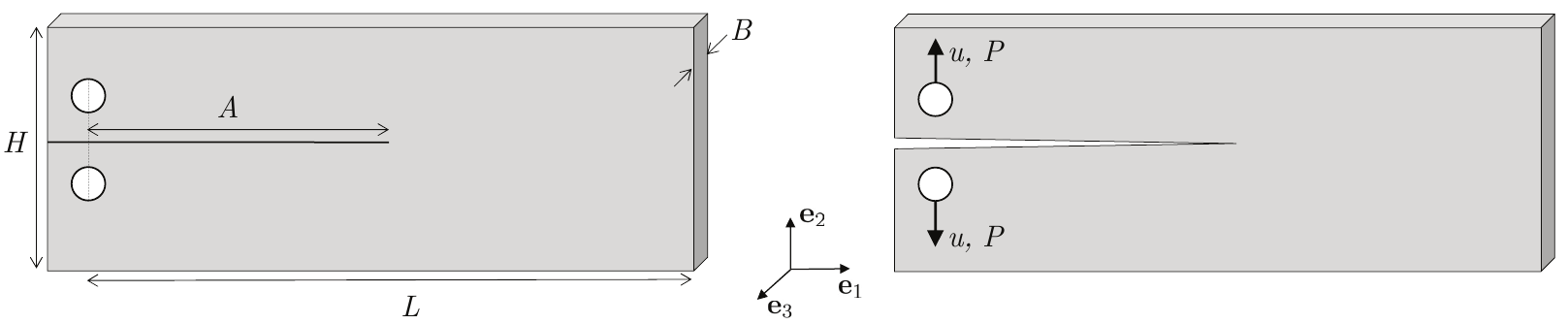}
\caption{\small Schematic of the double cantilever beam test. The specimen dimensions are $L=55$ mm, $H=20$ mm, $B=2.5$ mm, and $A=25$ mm. The radius of the pinholes is $B/8=0.3125$ mm and their centers are at a distance of $H/3=6.6667$ mm from each other and at a distance of $1.5$ mm from the left boundary of the specimen.}\label{Fig18}
\end{figure}

The eighth challenge problem is depicted in Fig.~\ref{Fig18}. It involves a prismatic bar of length $L=55$ mm and rectangular cross section of height $H=20$ mm and thickness $B=2.5$ mm, that contains a pre-existing edge crack of initial length $A=25$ mm. These dimensions are consistent with some of those found in the classical and recent experimental literature \citep{Gilman1960,Aaldenberg2022}. They have been suitably selected so that they can be used for both the hard soda-lime glass and the soft PU elastomer of interest in this work. The arms formed by the pre-existing cracks are pulled apart by a displacement $u$, applied through pinholes, in the direction $\bfe_2$ indicated in the figure. The resultant force is denoted by $P$. 

As the displacement $u$ is increased, the arms of the specimen bend like cantilever beams (hence the name of the test) until a critical value of $u$ is reached, say $u_{cr}$, at which point the Griffith criticality condition (\ref{Griffith-Cond-Propa}) is satisfied and the crack starts to grow straight ahead, in the direction $\bfe_1$ indicated in the figure. Further increase in $u$ leads to the continuous satisfaction of the Griffith criticality condition (\ref{Griffith-Cond-Propa}) and, by the same token, the continuous and stable growth of the crack in the $\bfe_1$ direction. To be viable, a computational model of fracture must be able to accurately predict these results.

For the case of double cantilever beam specimens of rectangular cross section, for which the boundary ahead of the crack is sufficiently far away, made of isotropic linear elastic brittle materials, \cite{Gross66} and \cite{Moses1968} employed a boundary collocation method to estimate in closed form the global force-displacement ($P$ vs. $u$) response and the derivative $-\partial\mathcal{P}/\partial\Gamma$. When used in conjunction with the Griffith criticality condition (\ref{Griffith-Cond-Propa}), the result for the global force-displacement can be written as
\begin{equation}
P=\left\{\begin{array}{ll}
\dfrac{B H^{3+\alpha_1}\mu E}{32 H^{\alpha_1}\mu A^3+3\times 2^{3+\alpha_1}\alpha_2 H^2\mu A^{1+\alpha_1}+6 \alpha_3 H^{2+\alpha_1}E A}u &{\rm if}\;0\leq u< u_{cr}\vspace{0.2cm}\\
\dfrac{B H^{3+\alpha_1}\mu E}{32  H^{\alpha_1}\mu a^3+3\times 2^{3+\alpha_1}\alpha_2 H^2\mu a^{1+\alpha_1}+6  \alpha_3 H^{2+\alpha_1}E a}u  &{\rm if}\;u_{cr}\leq u\end{array}\right.\label{P-u-DCB}
\end{equation}
in terms of the evolving length $a$ of the crack, given by
\begin{equation}
a=\left\{\begin{array}{ll}
A &{\rm if}\;0\leq u< u_{cr}\vspace{0.2cm}\\
\textrm{Root closest to } A \textrm{ of the non-linear algebraic equation}\,g(a;u)=0  &{\rm if}\;u_{cr}\leq u\end{array}\right.,\label{a-DCB}
\end{equation}
where
\begin{align*}
g(a;u)=\dfrac{2G_c a^2}{H^3\mu}-\dfrac{3H^{\alpha_1}E\left(16 H^{\alpha_1}\mu a^2+2^{2+\alpha_1}(1+\alpha_1)\alpha_2 H^2 \mu a^{\alpha_1}+\alpha_3 H^{2+\alpha_1}E\right)u^2}{\left(16 H^{\alpha_1}\mu a^2+3\times 2^{2+\alpha_1}\alpha_2 H^2\mu a^{\alpha_1}+3 \alpha_3 H^{2+\alpha_1}E\right)^2}.
\end{align*}
In these expressions, $\alpha_1$, $\alpha_2$, $\alpha_3$ are constants that exhibit small variations depending on how the boundary conditions are precisely applied. For the boundary conditions considered here, FE solutions indicate that $\alpha_1=1.03$, $\alpha_2=0.324$, and $\alpha_3=2.277$; cf. Table II in \citep{Moses1968} where $\alpha_1=n$, $\alpha_2=c$, and $\alpha_3=\alpha$. No analogous estimates to (\ref{P-u-DCB}) and (\ref{a-DCB}) have been worked out for specimens made of non-linear elastic brittle materials. Nevertheless, standard FE calculations allow to compute the corresponding results numerically. In the analysis of the double cantilever beam made of the PU elastomer presented here, we do just that.

\begin{figure}[t!]
\centering
\centering\includegraphics[width=0.99\linewidth]{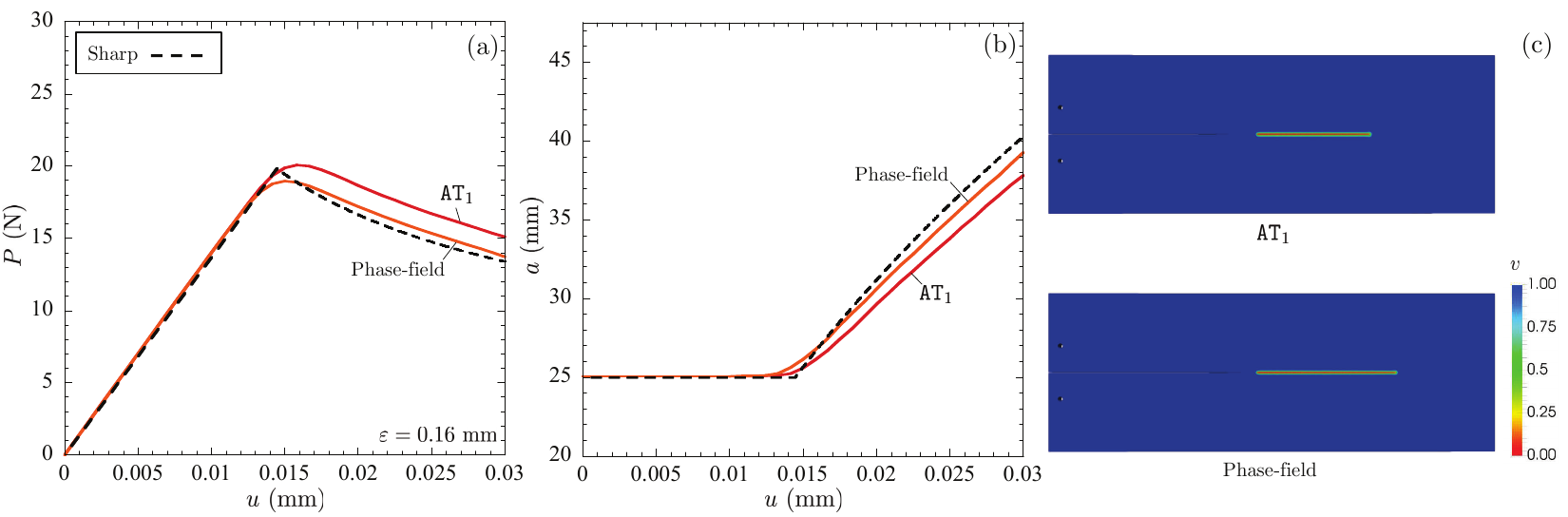}
\caption{\small Double cantilever beam test on the soda-lime glass. Comparisons between the exact results (\ref{P-u-DCB}) and (\ref{a-DCB}), for (a) the force-displacement response and (b) the evolution of crack length, and the predictions by the \texttt{AT}$_1$ and phase-field models for regularization length $\varepsilon=0.16$ mm. (c) Contour plots of the phase field $v$ over the deformed configuration of the specimen at $u=0.03$ mm, as predicted by the \texttt{AT}$_1$ and phase-field models.}\label{Fig19}
\end{figure}
\begin{figure}[t!]
\centering
\centering\includegraphics[width=0.99\linewidth]{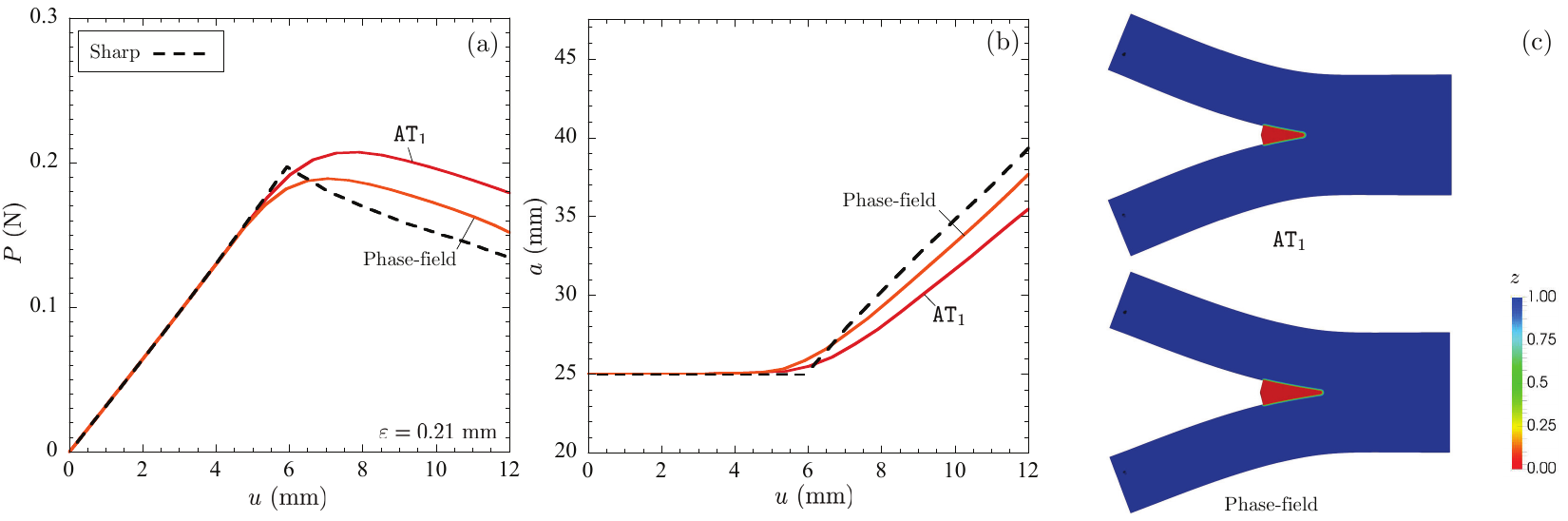}
\caption{\small Double cantilever beam test on the PU elastomer. Comparisons between the exact results computed from full-field FE simulations, for (a) the force-displacement response and (b) the evolution of crack length, and the predictions by the \texttt{AT}$_1$ and phase-field models for regularization length $\varepsilon=0.21$ mm. (c) Contour plots of the phase field $z$ over the deformed configuration of the specimen at $u=10$ mm, as predicted by the \texttt{AT}$_1$ and phase-field models.}\label{Fig20}
\end{figure}

To reduce computational cost in the \texttt{AT}$_1$ and phase-field simulations of the double cantilever beam test, we exploit symmetry and simulate only a quarter of the specimen. The FE mesh used for the simulations is unstructured and refined to an element size of $h=0.05$ mm ahead of the crack front, where the propagation of the crack is expected to take place. Figures \ref{Fig19}(a) and \ref{Fig20}(a) compare the force-displacement response ($P$ vs. $u$) of the specimen as predicted by the \texttt{AT}$_1$ and phase-field models with the corresponding exact results for the soda-lime glass and PU elastomer, respectively. Figures \ref{Fig19}(b) and \ref{Fig20}(b) provide the associated comparisons for the evolving length $a$ of the crack as a function of the applied displacement $u$. Finally,  Figs. \ref{Fig19}(c) and \ref{Fig20}(c) present contour plots of the phase fields, $v$ and $z$, over the deformed configuration of the specimen after the crack has undergone significant propagation, at $u=0.03$ mm in Fig.~\ref{Fig19}(c) and at $u=10$ mm in Fig.~\ref{Fig20}(c), as predicted by the \texttt{AT}$_1$ and phase-field models. The phase-field results for soda-lime glass pertain to the value $\varepsilon=0.16$ mm for the regularization length, while those for the PU elastomer pertain to $\varepsilon=0.21$ mm. These are the values for which the \texttt{AT}$_1$ model was fitted to describe the uniaxial tension test.

Consistent with all previous findings on fracture nucleation in Sections \ref{Sec: Strength Nucleation}, \ref{Sec: Griffith Nucleation}, and \ref{Sec: Mediation Nucleation} above, Figs.~\ref{Fig19} and \ref{Fig20} show that the predictions from the phase-field model are also in good agreement with fracture propagation in the double cantilever beam test, provided that the regularization length is sufficiently small. The figures also show that the \texttt{AT}$_1$  model's predictions happen to be in fair, but less accurate, agreement with the exact results.

\subsection{The trousers test}

\begin{figure}[H]
\centering
\centering\includegraphics[width=0.8\linewidth]{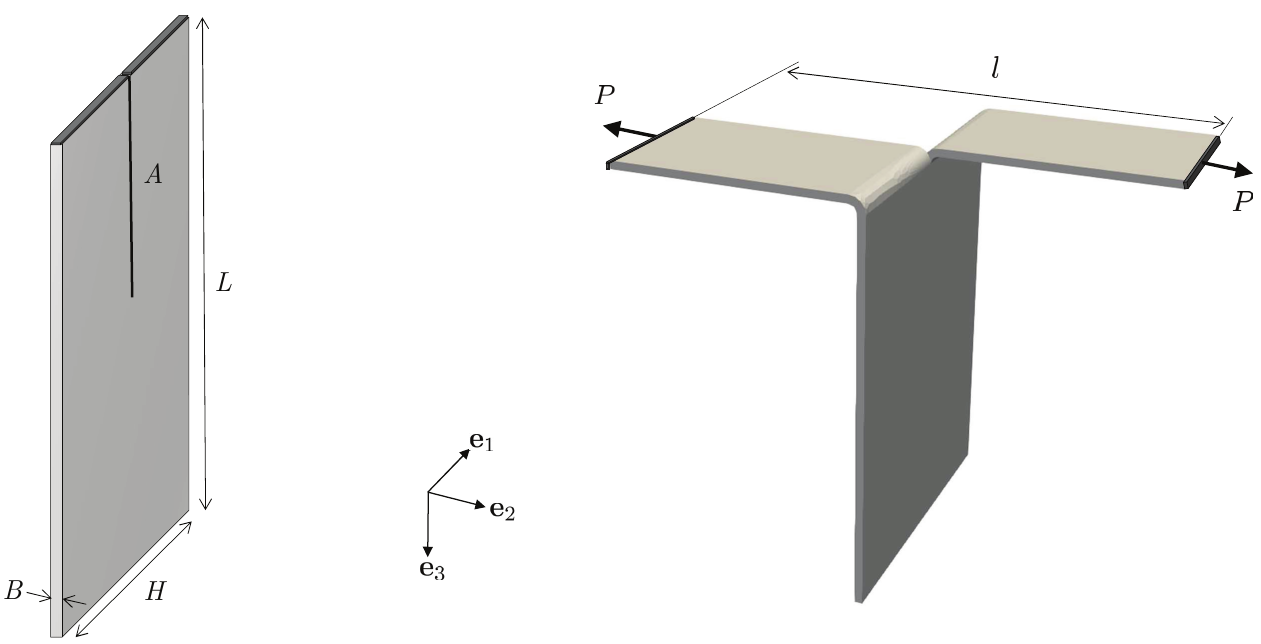}
\caption{\small Schematic of the trousers test. The specimen dimensions are $L=100$ mm, $H=40$ mm, $B=1$ mm, and $A=50$ mm.}\label{Fig21}
\end{figure}

The ninth and final challenge problem is the trousers test depicted in Fig.~\ref{Fig21}. The specimen comprises a sheet of length $L=100$ mm, width $H=40$ mm, and thickness $B=1$ mm that contains a pre-existing edge crack of initial length $A=50$ mm. These dimensions are consistent with those utilized by \cite{Greensmith55} in their pioneering trousers tests on rubber, as well as in the \cite{ASTM}. Assuming that the material is not exceedingly stiff so that the sheet can be bent, the specimen is loaded by first bending its two legs in opposite directions to bring them into the same plane. Subsequently, their bottom ends are held firmly by stiff grips, which are then pulled apart, in the direction $\bfe_2$ indicated in the figure. The pulling is done by applying a separation $l$ between the grips. The resultant force is denoted by $P$. 

\begin{figure}[t!]
\centering
\centering\includegraphics[width=0.98\linewidth]{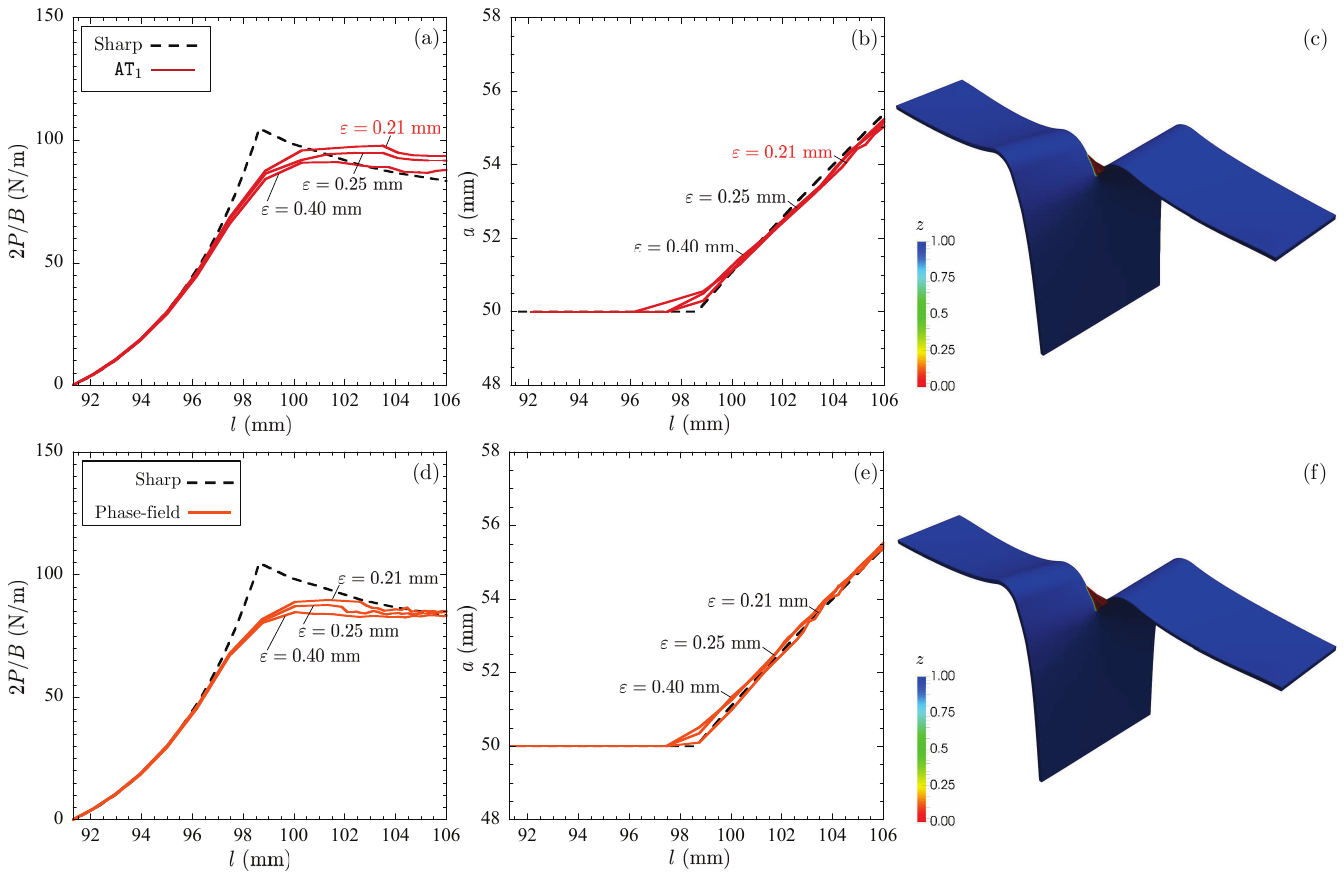}
\caption{\small Trousers test on the PU elastomer. Comparisons between the exact results computed from full-field FE simulations, for (a,d) the normalized-force-grip-separation response and (b,e) the evolution of crack length, and the predictions by (a,b) the \texttt{AT}$_1$ and (d,e) phase-field models for regularization lengths $\varepsilon=0.21, 0.25,$ and $0.40$ mm. (c,f) Contour plots of the phase field $z$ over the deformed configuration of the specimen at $l=106$ mm, as predicted by (c) the \texttt{AT}$_1$ and (f) phase-field models for $\varepsilon=0.21$ mm.}\label{Fig22}
\end{figure}

As the separation $l$ between the grips is increased, the legs of the specimen stretch, bend, and twist elastically until a critical value of $l$ is reached, say $l_{cr}$, at which point the Griffith criticality condition (\ref{Griffith-Cond-Propa}) is satisfied and the crack starts to grow in a self-similar manner, in the direction $\bfe_3$ indicated in the figure. Further increase in $l$ leads to a continuous satisfaction of the Griffith criticality condition (\ref{Griffith-Cond-Propa}) and, by the same token, a roughly continuous and stable growth of the crack in the $\bfe_3$ direction. For a computational model of fracture to be viable, it must be able to accurately predict these results.

In a recent contribution \citep{KLP25}, a full-field analysis has shown that the classical Rivlin-Thomas approximation ($-\partial\mathcal{P}/\partial\Gamma=2P/B+corrections$) that has been routinely used to estimate the derivative $-\partial\mathcal{P}/\partial\Gamma$ in (\ref{Griffith-Cond-Propa}) for trousers tests can be substantially inaccurate. Nevertheless, its computation via FE is straightforward. The results presented below for the PU elastomer are computed via the FE approach detailed in \citep{KLP25}. No results for soda-lime glass are included since this material is too stiff to deform and fracture in a self-similar manner for the dimensions of the specimen considered here.\footnote{In principle, glass specimens with thicknesses on the order of $B\sim 0.01$ mm (analogous to that of aluminum foil) could be utilized. Nevertheless, existing fabrication processes do not yet appear capable of attaining such small thicknesses.}

Figure \ref{Fig22} compares the normalized force $2P/B$ and evolving crack length $a$ predicted by the \texttt{AT}$_1$ and phase-field models with the corresponding exact results computed from full-field FE simulations for the PU elastomer. The results are shown as a function of the applied separation $l$ between the grips for three values of the regularization length $\varepsilon$. Figure \ref{Fig22} also presents contour plots of the phase field $z$ over the deformed configuration of the specimen after the crack has undergone significant propagation, at $l=106$ mm, as predicted by the \texttt{AT}$_1$ and phase-field models for $\varepsilon=0.21$ mm.

Consistent once more with all previous findings on fracture nucleation in Sections \ref{Sec: Strength Nucleation}, \ref{Sec: Griffith Nucleation}, and \ref{Sec: Mediation Nucleation} above, as well as those of Mode I fracture propagation in the preceding subsection, the results in Fig.~\ref{Fig22} show that the phase-field model accurately predicts Mode III fracture propagation in the trousers test, provided that the regularization length $\varepsilon$ is sufficiently small. Figure \ref{Fig22} also indicates that the predictions generated by the \texttt{AT}$_1$ model show fair, though less accurate, agreement with the exact results.

\section{Final comments}\label{Sec: Final comments}

Computational models of fracture are essential tools. They help us predict where and when structures and devices might fail, which is crucial for making them safe and reliable. They are also essential for thorough forensic investigations of past failures. In addition, they have the potential to empower scientists and engineers to answer outstanding fundamental questions and design new materials with enhanced fracture resistance. These models are also wanted for replacing and expanding on costly, full-scale physical testing, making them an invaluable tool for achieving a deeper understanding of material behavior. Given their critical importance and the increasingly complex class of problems on which they are being brought to bear --- often involving coupled physical phenomena --- interpreting results based on models that are insufficiently descriptive of reality is simply untenable.  

\begin{figure}[b!]
\centering
\centering\includegraphics[width=0.86\linewidth]{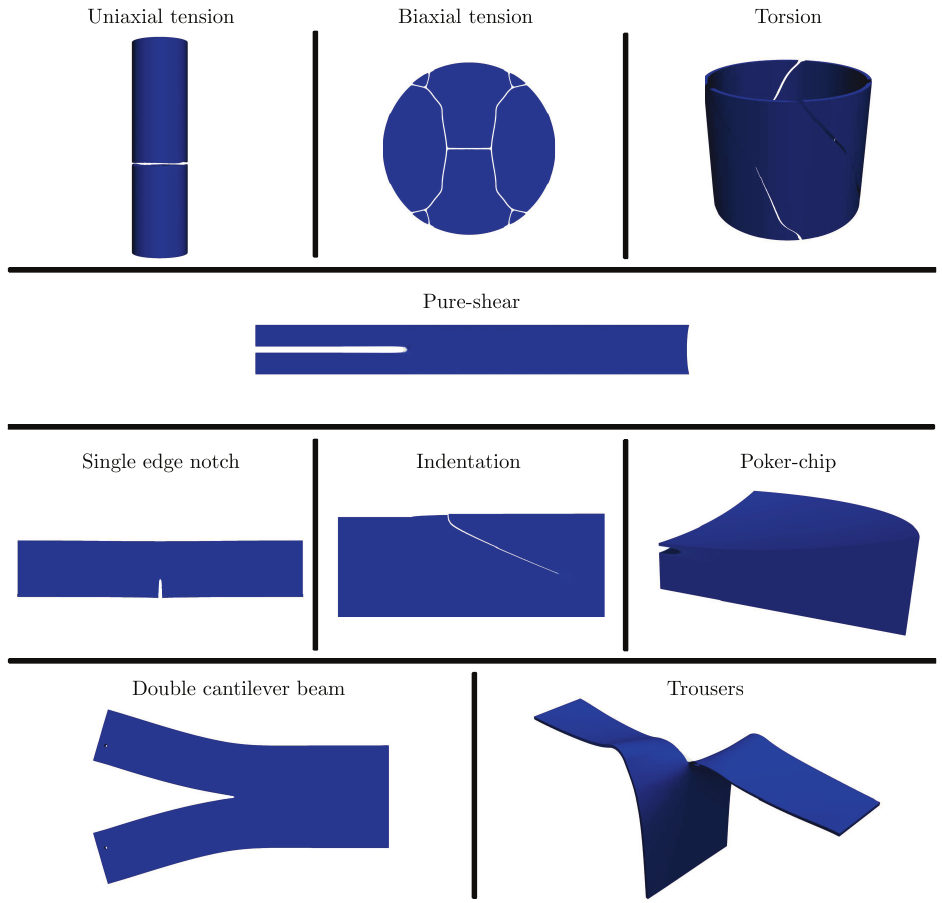}
\caption{\small Pictorial summary of the nine challenge problems illustrating the patterns of nucleated and propagated cracks in the soda-lime glass and the PU elastomer utilized in this work as representatives of hard and soft elastic brittle materials.}\label{Fig23}
\end{figure}

With this big picture in mind, the nine challenge problems presented in this paper aim at establishing a minimum standard --- within the simplest setting of isotropic elastic brittle materials subjected to quasi-static mechanical loads --- that any computational model of fracture ought to pass, if it is to potentially describe fracture nucleation and propagation in general. Not passing one test makes a model \emph{not} viable.\footnote{Some computational models rely on a set of ``free parameters'' that can be tuned to varying degrees in order to better fit experimental observations.  Since the material properties do not change between the nine tests, any computational model that requires  such parameters to be changed from one test to the next cannot, in fact, describe fracture nucleation and propagation in general.}  Passing all nine tests makes a model viable, but not necessarily foolproof. Additional direct quantitative comparisons with experiments should be performed to further validate the model. In other words, passing the nine challenge problems is a necessary but not sufficient condition for a computational model of fracture to be descriptive and predictive.

The representative results on a hard soda-lime glass and a soft PU elastomer presented throughout this work --- summarized pictorially in Fig.~\ref{Fig23} --- have served to illustrate the use of the challenge problems. As expounded at length in several recent works \citep{KLP20,KBFLP20,LPDFL25,KDLP25}, these results have also served to illustrate that only models that account for the elastic energy density $W(\bfE)$, or $\psi(\bfF)$, the strength surface $\mathcal{F}(\bfS)=0$, and the toughness $G_c$ in their entirety as three independent macroscopic material properties can possibly pass all nine challenge problems. This fundamental requirement rules out not only the \texttt{AT}$_1$ model used in this work for demonstration purposes, but also any existing classical variational or cohesive-type phase-field model, with or without energy splits, as well as any existing peridynamics model. Regardless, we encourage practitioners of such formulations and other variants to run their models through the circles and examine their findings.

With the objective of making the challenge problems as accessible as possible, FE meshes for each one of them have been made available on GitHub. We also note that 2D (axisymmetric and plane-stress) versions of the uniaxial tension test, biaxial tension test, pure-shear fracture test, single edge notch test, poker-chip test, and double cantilever beam test instead of the 3D versions examined here could be utilized to reduce computational cost, at the expense of not thoroughly probing the behavior of the model at hand. The torsion and trousers tests do not admit 2D simplifications and must be solved fully in 3D.

\section*{Acknowledgements}

This work was supported by the National Science Foundation through the Grants DMS--2308169, CMMI--2132528, CMMI--2132551. This support is gratefully acknowledged.

\section*{Appendix A. The phase-field formulation of \cite*{KFLP18}}

In this Appendix, we briefly recall the phase-field model introduced by \cite*{KFLP18}. For simplicity of exposition, consistent with the choices in the main body of the text of the soda-lime glass and the PU elastomer as representative hard and soft materials, we present the model for the basic case of elastic brittle materials with Neo-Hookean elasticity and its specialization to linear elasticity in the limit of small deformations. For a complete account, including mathematical well-posedness, thermodynamic consistency, FE schemes to solve the resulting governing equations, and validation with experiments the interested reader is referred to \cite{KFLP18,KRLP18,KLP20,KBFLP20,KLP21,KRLP22,KLDLP24,LDLP24}; and \cite{KKLP24}.

\paragraph{Initial configuration, kinematics, and boundary conditions} Consider a generic specimen that, initially, at time $t=0$, occupies the open bounded domain $\Omega$. We denote its boundary by $\partial \Omega$, its outward unit normal by $\bfN$, and identify material points by their initial position vector $\bfX\in \Omega$. The specimen is subjected a deformation $\overline{\bfy}(\bfX,t)$ on a part $\partial \Omega_{\mathcal{D}}$ of the boundary, and a surface force (per unit undeformed area) $\overline{\bfs}(\bfX,t)$ on the complementary part $\partial \Omega_{\mathcal{N}}=\partial\Omega\setminus\partial \Omega_{\mathcal{D}}$; for simplicity, body forces are considered to be absent. In response to these boundary conditions, the position vector $\bfX$ of a material point in the specimen will move to a new position specified by 
\begin{equation}
\bfx=\bfy(\bfX,t)=\bfX+\bfu(\bfX,t),\label{def-mapping}
\end{equation}
where $\bfy(\bfX,t)$ and $\bfu(\bfX,t)$ are the deformation and displacement fields. We write the associated deformation gradient at $\bfX$ and $t$ as $\bfF(\bfX,t)=\nabla\bfy(\bfX,t)=\bfI+\nabla\bfu(\bfX,t)$. In addition to the deformation (\ref{def-mapping}), the boundary conditions may result in the nucleation and subsequent propagation of cracks in the specimen. We describe such cracks in a regularized fashion via the phase field
\begin{equation*}
z=z(\bfX,t)
\end{equation*}
taking values in the range $[0, 1]$. The value $z=1$ identifies the intact regions of the material and $z=0$ the regions that have been fractured, while the transition from $z=1$ to $z=0$ is set to occur smoothly over regions of small thickness of regularization length scale $\varepsilon>0$.

\paragraph{Constitutive behavior} As noted above, the specimen is taken to be made of a homogeneous, isotropic, elastic brittle material whose elastic behavior is characterized by the Neo-Hookean elastic energy density (\ref{W-NH}). It follows that the first Piola-Kirchhoff stress at any material point $\bfX\in\Omega$ and time $t\in[0,T]$ is given by
\begin{equation}
\bfS(\bfX,t)=\dfrac{\partial \psi}{\partial \bfF}(\bfF)=\mu(\bfF-\bfF^{-T})+\Lambda(\det\bfF-1)(\det\bfF)\bfF^{-T}.\label{S-NH}
\end{equation}
Moreover, the strength of the material is characterized by the Drucker-Prager strength surface (\ref{DP}), where we recall that $\mathcal{I}_1=s_1+s_2+s_3$ and $\mathcal{J}_2=1/3(s_1^2+s_2^2+s_3^2-s_1 s_2-s_1 s_3-s_2s_3)$ stand for the first and second principal invariants of the Biot stress tensor $\bfS^{(1)}=(\bfS^T\bfR+\bfR^T\bfS)/2$, with $s_1$, $s_2$, $s_3$ denoting the principal Biot stresses and $\bfR$ the rigid rotation from the polar decomposition $\bfF=\bfR\bfU$, while the material constants $s_{\texttt{ts}}>0$ and $s_{\texttt{hs}}>0$ stand for the uniaxial tensile and hydrostatic strengths of the material. That is, they denote the critical nominal stress values at which fracture nucleates under uniform states of monotonically increased uniaxial tension $\bfS={\rm diag}(s>0,0,0)$ and hydrostatic stress $\bfS={\rm diag}(s>0,s>0,s>0)$, respectively. Finally, the critical energy release rate of the material is characterized by the scalar (\ref{Gc}).

\paragraph{Governing equations} According to the phase-field fracture formulation put forth by \citet*{KFLP18}, the deformation field $\bfy_k(\bfX)=\bfy(\bfX,t_k)$ and phase field $z_k(\bfX)=z(\bfX,t_k)$ at any material point $\bfX \in \overline{\Omega}=\Omega\cup\partial\Omega$ and at any given discrete time $\{t_{k}\}_{k=0,1,\ldots,M}$, with $t_{0} = 0$ and $t_{M} = T$, are determined by the system of coupled partial differential equations\footnote{The Neumann boundary condition (\ref{BVP-z-theory})$_2$ may be replaced by the Dirichlet boundary condition $z_k=0$ at the front of any pre-existing crack and sharp corner that the boundary $\partial\mathrm{\Omega}$ of the specimen may feature. The simulations for the pure-shear test, the double cantilever beam test, and the trousers test in the main body of the text are carried out with this Dirichlet boundary condition. On the other hand, the simulations for the single edge notch test are carried out with the Neumann boundary condition.}
\begin{equation}
\left\{\begin{array}{ll}
{\rm Div}\left[z_{k}^2 (2\mu(\nabla \bfy_{k}-\nabla \bfy_{k}^{-T})+\Lambda(J_k-1)J_k\nabla \bfy_{k}^{-T})\right]=\textbf{0},& \,\bfX\in\Omega\vspace{0.2cm}\\
\bfy_k(\bfX)=\overline{\bfy}(\bfX,t_k), & \; \bfX\in\partial\Omega_{\mathcal{D}}\vspace{0.2cm}\\
\left(z_{k}^2 (2\mu(\nabla \bfy_{k}-\nabla \bfy_{k}^{-T})+\Lambda(J_k-1)J_k\nabla \bfy_{k}^{-T})\right)\bfN=\overline{\bfs}(\bfX,t_k),& \; \bfX\in\partial\Omega_{\mathcal{N}}
\end{array}\right. \label{BVP-y-theory}
\end{equation}
and
\begin{equation}
\left\{\begin{array}{ll}
\hspace{-0.15cm} \varepsilon\, \delta^\varepsilon G_c \triangle z_k=\dfrac{8}{3}z_{k} \psi(\nabla\bfy_k)-\dfrac{4}{3}c_\texttt{e}(\bfX,t_{k})-\dfrac{\delta^\varepsilon G_c}{2\varepsilon}+\dfrac{8}{3\,\zeta} \, p(z_{k-1},z_k),& \bfX\in \Omega
\vspace{0.2cm}\\
\hspace{-0.15cm}\nabla z_k\cdot\bfN=0,&  \bfX\in \partial\Omega
\end{array}\right. \label{BVP-z-theory}
\end{equation}
with $p(z_{k-1},z_k)=|z_{k-1}-z_k|-(z_{k-1}-z_k)-|z_k|+z_k$. In these equations, $\nabla \bfy_k(\bfX)=\nabla \bfy(\bfX,t_k)$, $J_k=\det\nabla\bfy_k$, $\nabla z_k(\bfX)=\nabla z(\bfX,t_k)$, $\triangle z_k(\bfX)=\triangle z(\bfX,t_k)$, $\zeta$ is a penalty parameter\footnote{The penalty function $p(z_{k-1},z_k)$ and penalty parameter $\zeta$ in (\ref{BVP-z-theory}) enforce that the phase field remains in the physically admissible range $0\leq z\leq 1$ and that fracture is irreversible. These requirements can be enforced by means of different strategies \citep{Wick15}, the penalty approach spelled out here being one of them. In the RACCOON implementation, for example, these inequalities are enforced using the primal-dual active set strategy.}  such that\footnote{Typically, it suffices to set $\zeta^{-1}=10^4 \delta^\varepsilon G_c/(2\varepsilon)$.} $\zeta^{-1}\gg \delta^\varepsilon G_c/(2\varepsilon)$, and, making use of the constitutive prescription\footnote{The constitutive prescription for $c_{\texttt{e}}$ depends on the particular form of the strength surface $\mathcal{F}(\bfS)=0$ of the material. For the case of the Drucker-Prager strength surface (\ref{DP}) of interest here, it is given by (\ref{ce}). For other strength surfaces, corresponding prescriptions for $c_{\texttt{e}}$ can be constructed by following the blueprint outlined by \cite{KLP20} and \cite{KBFLP20}.} in \citep{KKLP24}, 
\begin{align}
c_{\texttt{e}}(\bfX,t)&=z^2\beta_2\sqrt{\mathcal{J}_2}+z^2\beta_1\mathcal{I}_1+z\left(1-\dfrac{\sqrt{\mathcal{I}^2_1}}{\mathcal{I}_1}\right)\psi(\bfF)\label{ce}
\end{align}
with 
\begin{equation}
\left\{\hspace{-0.15cm}\begin{array}{l}
\beta_1=-\dfrac{1}{\shs}\delta^\varepsilon\dfrac{G_c}{8\varepsilon}+\dfrac{2\psi_{\texttt{hs}}}{3\shs}\vspace{0.2cm}\\
\beta_2=-\dfrac{\sqrt{3}(3\shs-\sts)}{\shs\sts}\delta^\varepsilon\dfrac{G_c}{8\varepsilon}-
\dfrac{2\psi_{\texttt{hs}}}{\sqrt{3}\shs}+\dfrac{2\sqrt{3}\psi_{\texttt{ts}}}{\sts}
\end{array}\right.\quad {\rm and}\quad \delta^\varepsilon=\left(\dfrac{\sts+(1+2\sqrt{3})\,\shs}{(8+3\sqrt{3})\,\shs}\right)\dfrac{3 G_c}{16\psi_{\texttt{ts}}\varepsilon}+\dfrac{2}{5}. \label{deltaeps}
\end{equation}
In these last expressions, $\psi_{\texttt{ts}}$ and $\psi_{\texttt{hs}}$ stand for the values of the stored-energy function (\ref{W-NH}) along uniform uniaxial tension and hydrostatic stress states at which the strength surface (\ref{DP}) is violated:
\begin{equation*}
\psi_{\texttt{ts}}=\dfrac{\mu}{2}(\l^2_{\texttt{ts}}+2\l^2_l-3)-\mu\ln(\l_{\texttt{ts}}\l_l^2)+\dfrac{\Lambda}{2}(\l_{\texttt{ts}}\l_l^2-1)^2\quad {\rm and}\quad \psi_{\texttt{hs}}=\dfrac{\mu}{2}(3\l^2_{\texttt{hs}}-3)-\mu\ln(\l^3_{\texttt{hs}})+\dfrac{\Lambda}{2}(\l^3_{\texttt{hs}}-1)^2,
\end{equation*}
where the pair of stretches ($\l_{\texttt{ts}}, \l_l$) and the stretch $\l_{\texttt{hs}}$ are defined implicitly as the roots closest to $(1,1)$ and $1$ of the non-linear algebraic equations
\begin{align*}
&\left\{\begin{array}{l}
s_{\texttt{ts}}=\mu(\l_{\texttt{ts}}-\l_{\texttt{ts}}^{-1})+\Lambda(\l_{\texttt{ts}}\l_l^4-\l_l^2)\\[10pt]
0=\mu(\l_l-\l_l^{-1})+\Lambda\l_{\texttt{ts}}\l_l(\l_{\texttt{ts}}\l_l^2-1)\end{array}\right. \quad {\rm and}\quad s_{\texttt{hs}}=\mu(\lambda_{\texttt{hs}}-\lambda_{\texttt{hs}}^{-1})+\Lambda\lambda_{\texttt{hs}}^2(\lambda_{\texttt{hs}}^3-1),  
\end{align*}
respectively. For the PU elastomer with the material constants listed in Table \ref{Table2},  $\l_{\texttt{ts}}=1.2342$, $\l_l=0.9007$, $\lambda_{\texttt{hs}}=1.0038$, and hence $\psi_{\texttt{ts}}=37.31$ kPa and $\psi_{\texttt{hs}}=5.718$ kPa.

\begin{remark}
\emph{As noted throughout the main body of the text, the strength of a material is inherently stochastic. This is because the strength at any given macroscopic material point depends on the varying nature of the underlying defects where fracture originates. Consequently, the strength constants $\sts$ and $\shs$ in the equations above should be considered as stochastic material constants, and not as deterministic values. In all the phase-field simulations presented in the main body of the text, $\sts$ is assigned random values within $\pm 5\%$ of those listed in Tables \ref{Table1} and \ref{Table2}. These random values are applied across random subdomains of size $5\varepsilon$.}
\end{remark}

\begin{remark}
\emph{When using the FE method to solve the governing equations (\ref{BVP-y-theory})-(\ref{BVP-z-theory}), meshes of small enough element size $\texttt{h}$ ought to be used so as to appropriately resolve the spatial variations of the phase field $z_k$ over lengths of order $\varepsilon$. Nevertheless, an error is incurred that scales with $\texttt{h}$. It is possible to include a correction in the formula (\ref{deltaeps})$_2$ for $\delta^\varepsilon$ so that the FE solutions of equations (\ref{BVP-y-theory})-(\ref{BVP-z-theory}) are consistent with the actual value $G_c$ of the critical energy release rate of the material. For first-order FEs of size $\texttt{h}$, the formula for $\delta^\varepsilon$ with the correction reads \citep{KKLP24}
\begin{equation*}
\delta^\varepsilon=\left(1+\dfrac{3}{8}\dfrac{\texttt{h}}{\varepsilon}\right)^{-2}\left(\dfrac{\sts+(1+2\sqrt{3})\,\shs}{(8+3\sqrt{3})\,\shs}\right)\dfrac{3 G_c}{16\psi_{\texttt{ts}}\varepsilon}+\left(1+\dfrac{3}{8}\dfrac{\texttt{h}}{\varepsilon}\right)^{-1}\dfrac{2}{5}.
\label{delta-eps-final-h}
\end{equation*}
}
\end{remark}

\paragraph{Governing equations: The limit of small deformations} In the limit of small deformations, the elasticity of the material reduces to isotropic linear elasticity. In particular, to leading order, the elastic energy density (\ref{W-NH}) and stress-deformation relation (\ref{S-NH})  reduce to (\ref{W-Lin}) and
\begin{equation*}
\bfS(\bfX,t)=\dfrac{\partial W}{\partial \bfE}(\bfE)=2\mu\,\bfE+\Lambda ({\rm tr}\,\bfE)\bfI,
\end{equation*}
respectively. Making use of the displacement field $\bfu(\bfX,t)=\bfy(\bfX,t)-\bfX$ instead of the deformation field $\bfy(\bfX,t)$ and rewriting the phase-field as $v(\bfX,t)$ instead of $z(\bfX,t)$ for further clarity of notation, it follows that the governing equations (\ref{BVP-y-theory})-(\ref{BVP-z-theory}) reduce to 
\begin{equation}
\left\{\begin{array}{ll}
\hspace{-0.15cm} {\rm Div}\left[v_{k}^2 \left(2\mu\,\bfE(\bfu_{k})+\Lambda({\rm tr}\,\bfE(\bfu_{k}))\bfI\right)\right]=\textbf{0},& \bfX\in\Omega\vspace{0.2cm}\\
\hspace{-0.15cm} \bfu_k(\bfX)=\overline{\bfu}(\bfX,t_k), & \bfX\in\partial\Omega_{\mathcal{D}}\vspace{0.2cm}\\
\hspace{-0.15cm}v_{k}^2 \left(2\mu\,\bfE(\bfu_{k})+\Lambda ({\rm tr}\,\bfE(\bfu_{k}))\bfI\right)\bfN=\overline{\bfs}(\bfX,t_k),&  \bfX\in\partial\Omega_{\mathcal{N}}
\end{array}\right. \label{BVP-u-theory}
\end{equation}
and
\begin{equation}
\left\{\begin{array}{ll}
\hspace{-0.15cm} \varepsilon\, \delta^\varepsilon G_c \triangle v_k=\dfrac{8}{3}v_{k} W(\bfE(\bfu_k))-\dfrac{4}{3}c_\texttt{e}(\bfX,t_{k})-\dfrac{\delta^\varepsilon G_c}{2\varepsilon}+\dfrac{8}{3\,\zeta} \, p(v_{k-1},v_k),& \bfX\in \Omega
\vspace{0.2cm}\\
\hspace{-0.15cm}\nabla v_k\cdot\bfN=0,&  \bfX\in \partial\Omega
\end{array}\right., \label{BVP-v-theory}
\end{equation}
where $\overline{\bfu}(\bfX,t)=\overline{\bfy}(\bfX,t)-\bfX$ and where
\begin{align*}
c_{\texttt{e}}(\bfX,t)&=v^2\beta_2\sqrt{\mathcal{J}_2}+v^2\beta_1\mathcal{I}_1+v\left(1-\dfrac{\sqrt{\mathcal{I}^2_1}}{\mathcal{I}_1}\right)W(\bfE(\bfu))
\end{align*}
with $\mathcal{I}_1=(2\mu+3\Lambda){\rm tr}\,\bfE$, $\mathcal{J}_2=2\mu^2{\rm tr}\,\bfE_D^2$, $\bfE_D=\bfE-\frac{1}{3}({\rm tr}\,\bfE)\bfI$, 
\begin{equation*}
\left\{\hspace{-0.15cm}\begin{array}{l}
\beta_1=-\dfrac{1}{\shs}\delta^\varepsilon\dfrac{G_c}{8\varepsilon}+\dfrac{2 W_{\texttt{hs}}}{3\shs}\vspace{0.2cm}\\
\beta_2=-\dfrac{\sqrt{3}(3\shs-\sts)}{\shs\sts}\delta^\varepsilon\dfrac{G_c}{8\varepsilon}-
\dfrac{2W_{\texttt{hs}}}{\sqrt{3}\shs}+\dfrac{2\sqrt{3}W_{\texttt{ts}}}{\sts}
\end{array}\right.,\quad \quad \delta^\varepsilon=\left(\dfrac{\sts+(1+2\sqrt{3})\,\shs}{(8+3\sqrt{3})\,\shs}\right)\dfrac{3 G_c}{16 W_{\texttt{ts}}\varepsilon}+\dfrac{2}{5},
\end{equation*}
$W_{\texttt{ts}}=\sts^2/(2E)$, and $W_{\texttt{hs}}=\shs^2/(2\kappa)$. Note that for the soda-lime glass with the material constants listed in Table \ref{Table1}, $W_{\texttt{ts}}=11.43$ kPa and $W_{\texttt{hs}}=9.27$ kPa.

\section*{Appendix B. The classical variational \texttt{AT}$_1$ phase-field model}

The strength surface $\mathcal{F}(\bfS)=0$ of the material enters the phase-field model (\ref{BVP-y-theory})-(\ref{BVP-z-theory}) via the driving force $c_{\texttt{e}}$ and the coefficient $\delta^\varepsilon$. When the presence of the strength surface is removed, 
\begin{equation*}
c_{\texttt{e}}=0\quad {\rm and}\quad \delta^\varepsilon=1,
\end{equation*}
the phase-field model (\ref{BVP-y-theory})-(\ref{BVP-z-theory}) reduces to the classical variational \texttt{AT}$_1$ phase-field model for a Neo-Hookean elastic material:
\begin{equation}
\left\{\begin{array}{ll}
{\rm Div}\left[z_{k}^2 (2\mu(\nabla \bfy_{k}-\nabla \bfy_{k}^{-T})+\Lambda(J_k-1)J_k\nabla \bfy_{k}^{-T})\right]=\textbf{0},& \,\bfX\in\Omega\vspace{0.2cm}\\
\bfy_k(\bfX)=\overline{\bfy}(\bfX,t_k), & \; \bfX\in\partial\Omega_{\mathcal{D}}\vspace{0.2cm}\\
\left(z_{k}^2 (2\mu(\nabla \bfy_{k}-\nabla \bfy_{k}^{-T})+\Lambda(J_k-1)J_k\nabla \bfy_{k}^{-T})\right)\bfN=\overline{\bfs}(\bfX,t_k),& \; \bfX\in\partial\Omega_{\mathcal{N}}
\end{array}\right. \label{BVP-y-theory-AT1}
\end{equation}
and
\begin{equation}
\left\{\begin{array}{ll}
\hspace{-0.15cm} \varepsilon\,  G_c \triangle z_k=\dfrac{8}{3}z_{k} \psi(\nabla\bfy_k)-\dfrac{G_c}{2\varepsilon}+\dfrac{8}{3\,\zeta} \, p(z_{k-1},z_k),& \bfX\in \Omega
\vspace{0.2cm}\\
\hspace{-0.15cm}\nabla z_k\cdot\bfN=0,&  \bfX\in \partial\Omega
\end{array}\right. . \label{BVP-z-theory-AT1}
\end{equation}

By the same token, when the strength surface $\mathcal{F}(\bfS)=0$ of the material is not accounted for, the phase-field model (\ref{BVP-u-theory})-(\ref{BVP-v-theory}) reduces to the classical variational \texttt{AT}$_1$ phase-field model for an isotropic linear elastic material:
\begin{equation}
\left\{\begin{array}{ll}
\hspace{-0.15cm} {\rm Div}\left[v_{k}^2 \left(2\mu\,\bfE(\bfu_{k})+\Lambda({\rm tr}\,\bfE(\bfu_{k}))\bfI\right)\right]=\textbf{0},& \bfX\in\Omega\vspace{0.2cm}\\
\hspace{-0.15cm} \bfu_k(\bfX)=\overline{\bfu}(\bfX,t_k), & \bfX\in\partial\Omega_{\mathcal{D}}\vspace{0.2cm}\\
\hspace{-0.15cm}v_{k}^2 \left(2\mu\,\bfE(\bfu_{k})+\Lambda ({\rm tr}\,\bfE(\bfu_{k}))\bfI\right)\bfN=\overline{\bfs}(\bfX,t_k),&  \bfX\in\partial\Omega_{\mathcal{N}}
\end{array}\right. \label{BVP-u-theory-AT1}
\end{equation}
and
\begin{equation}
\left\{\begin{array}{ll}
\hspace{-0.15cm} \varepsilon\, G_c \triangle v_k=\dfrac{8}{3}v_{k} W(\bfE(\bfu_k))-\dfrac{G_c}{2\varepsilon}+\dfrac{8}{3\,\zeta} \, p(v_{k-1},v_k),& \bfX\in \Omega
\vspace{0.2cm}\\
\hspace{-0.15cm}\nabla v_k\cdot\bfN=0,&  \bfX\in \partial\Omega
\end{array}\right. . \label{BVP-v-theory-AT1}
\end{equation}

\begin{remark}
\emph{When using first-order FEs of size $\texttt{h}$ to solve the governing equations (\ref{BVP-y-theory-AT1})-(\ref{BVP-z-theory-AT1}), or (\ref{BVP-u-theory-AT1})-(\ref{BVP-v-theory-AT1}), the correction 
\begin{equation*}
G_c\mapsto \left(1+\dfrac{3\texttt{h}}{8\varepsilon}\right)^{-1}G_c
\end{equation*}
may be used to account for the error incurred by the discretization \citep{Bourdin08}. The simulations for the pure-shear test, the single edge notch test, the double cantilever beam test, and the trousers test in the main body of the text are carried out making use of this correction.
}
\end{remark}

\bibliographystyle{elsarticle-harv}
\bibliography{References}

\end{document}